\documentclass[12pt,useAMS, usenatbib]{mn2e}
\usepackage{amsmath}
\usepackage{amssymb}
\usepackage{graphicx}
\usepackage{subfigure}
\usepackage{mathrsfs}
\newcommand{\be}{\begin{equation}} 
\newcommand{\ee}{\end{equation}}

\newcommand{\ba}{\begin{eqnarray}} 
\newcommand{\ea}{\end{eqnarray}}

\newcommand{\Max}{{\rm max}}

\newcommand{\In}{{\rm in}}
\newcommand{\Out}{{\rm out}}
\newcommand{\oct}{{\rm oct}}
\newcommand{\ain}{a_\In}
\newcommand{\ein}{e_\In}
\newcommand{\aout}{a_\Out}
\newcommand{\eout}{e_\Out}
\newcommand{\emax}{e_\Max}
\newcommand{\eobs}{e_{\rm obs}}
\newcommand{\tide}{\rm Tide}
\newcommand{\rtide}{R_{\tide}}

\newcommand{\hatLin}{{\bf \hat{L}}_\In}
\newcommand{\hatLout}{{\bf \hat{L}}_\Out}

\newcommand{\Lvec}{{\bf L}}
\newcommand{\evec}{{\bf e}}

\newcommand{\aouteff}{a_{\rm out,eff}}
\newcommand{\gr}{\rm GR}
\newcommand{\tk}{t_{\rm k}}

\newcommand{\mjup}{M_{\rm J}}
\newcommand{\msun}{M_{\odot}}
\newcommand{\rjup}{R_{\rm J}}

\def\go{\mathrel{\raise.3ex\hbox{$>$}\mkern-14mu
             \lower0.6ex\hbox{$\sim$}}}

\def\lo{\mathrel{\raise.3ex\hbox{$<$}\mkern-14mu
             \lower0.6ex\hbox{$\sim$}}}
\begin{document} 

\title[Eccentric Warm Jupiters from Exterior Companions]{Moderately Eccentric Warm Jupiters from Secular Interactions with Exterior Companions} 
\pagerange{\pageref{firstpage}--\pageref{lastpage}} \pubyear{2017}

\label{firstpage}

\author[K. R. Anderson \& D. Lai]{Kassandra R. Anderson$^{1}$\thanks{E-mail: kra46@cornell.edu} \& Dong Lai$^{1,2}$ \\ \\ $^{1}$Cornell Center for Astrophysics and Planetary Science, Department of Astronomy, Cornell University, Ithaca, NY 14853, USA \\ $^{2}$Institute for Theory and Computation, Harvard-Smithsonian Center for Astrophysics, Cambridge, MA 02138, USA}

\maketitle

\begin{abstract}
Recent studies have proposed that most warm Jupiters (WJs, giant planets with
semi-major axes in the range of 0.1-1 AU) probably form in-situ, or
arrive in their observed orbits through disk migration. However, both
in-situ formation and disk migration, in their simplest flavors,
predict WJs to be in low-eccentricity orbits, in contradiction with
many observed WJs that are moderately eccentric ($e=0.2-0.7$). This
paper examines the possibility that the WJ eccentricities are raised
by secular interactions with exterior giant planet companions, following in-situ formation or migration on a circular orbit.
Eccentricity growth may arise from an inclined companion (through
Lidov-Kozai cycles), or from an eccentric, nearly coplanar companion (through apsidal precession resonances).  We
quantify the necessary conditions (in terms of the eccentricity,
semi-major axis and inclination) for external perturbers of various
masses to raise the WJ eccentricity.  We also consider the sample of
eccentric WJs with detected outer companions, and for each system,
identify the range of mutual inclinations needed to generate the
observed eccentricity. For most systems, we find that relatively high
inclinations (at least $\sim 40^\circ$) are needed so that Lidov-Kozai
cycles are induced; the observed outer companions are typically not
sufficiently eccentric to generate the observed WJ eccentricity in a
low-inclination configuration.  The results of this paper place
constraints on possibly unseen external companions to eccentric WJs.
Observations that probe mutual inclinations of giant planet systems will help
clarify the origin of eccentric WJs and the role of external companions.

\end{abstract}


\section{Introduction}
Despite over twenty years of observations, the origins and dynamical histories of close-in ($\lesssim 1$ AU) giant planets remain elusive.  Hot Jupiter (HJ, giant planets with semi-majors axes $\lesssim 0.1$ AU)  formation continues to be a major topic in exoplanet research.   The general consensus is that HJs cannot form in their present locations, and must instead have migrated from farther out (although see \citealt{boley2016,batygin2016}), but whether there exists a dominant migration mechanism is unclear.  Proposed mechanisms include disk-migration, and various forms of ``high-eccentricity migration'' in which the planet's eccentricity is excited to a large value, leading to tidal dissipation during pericenter passages and orbital decay.  Warm Jupiters (WJs, with semi-major axes in the range $\sim 0.1 - 1$ AU) raise the same formation questions as HJs.  Proposed channels of WJ formation include disk migration, high-eccentricity migration, scatterings, and in-situ formation.  If multiple channels of WJ formation exist, whether one channel produces most of the observed WJs is of great interest but remains unknown.

Many WJs are moderately eccentric, with $e \sim 0.2 - 0.7$.  These eccentricities are difficult to explain with both in-situ formation and disk-driven migration.  High-eccentricity migration has therefore been proposed as a major formation mechanism for WJs.  If WJs are undergoing high-eccentricity migration, they must reach sufficiently small pericenter distances ($\lesssim 0.05$ AU) to experience tidal dissipation and orbital decay.  Most WJs are not sufficiently eccentric to achieve such small pericenter distances, but this issue can be circumvented if the planets are undergoing secular eccentricity oscillations induced by exterior companions, and are currently observed in a lower eccentricity phase.  The requirement that the minimum pericenter distance be small enough such that tidal decay may occur within the lifetime of the host star constrains the properties of the perturbers, requiring them to be sufficiently close and/or massive \citep{dong2014}.  

However, the proposal that most WJs reach their current orbits through high-eccentricity migration suffers from some problems.  \cite{antonini2016} find that most observed WJs with exterior planetary companions would not be stable if the WJ originated beyond $\sim 1$ AU and subsequently underwent high-eccentricity migration.  Furthermore, population synthesis studies of HJ formation by various high-eccentricity migration mechanisms typically yield very low fractions of planets at WJ distances (\citealt{petrovich2015}, 2015b, \citealt{kra2016}, \citealt{hamers2017}, \citealt{hamers2017b}; although see \citealt{dawson2014}, \citealt{petrovich2016}).  For example, studies of high-eccentricity migration due to Lidov-Kozai oscillations from stellar perturbers \citep{petrovich2015, kra2016} produce HJs at rates of a few percent, but essentially no WJs.  This arises because, for a stellar perturber at a distance of $\sim$ few hundred AU, once the planetary orbit shrinks to WJ distances, eccentricity oscillations have ceased due to general relativistic precession, and the eccentricity has frozen to very high value ($\sim 0.99$), after which the migration to HJ distances proceeds rapidly (see, e.g. Fig.~1 of \citealt{kra2016}).  \cite{hamers2017} find a similarly negligible amount of WJs compared to HJs for high-eccentricity migration due to secular chaos in systems of multiple giant planets.  Observations of giant planets paint a very different picture. Despite the existence of a ``period valley'' of giant planets with orbital periods of $10-20 $ days \citep[e.g.][]{udry2003, jones2003, santerne2016}, the total occurrence rate of WJs (with semi-major axes in the range 0.1AU-1AU) exceeds that of HJs \citep[$a<0.1$ AU  see][Fig.~8]{santerne2016}. We note that the ratio of WJs to HJs does depend somewhat on the definition of a WJ.  Taking the RV planets listed on exoplanets.org\footnote{accessed on August 22, 2017.} with $m \sin i > 0.5 \mjup$, we find that the WJ/HJ ratio is $\sim 3.9$.   If we adopt a more conservative definition of a WJ, with $0.1  {\rm AU} < a < 0.5  {\rm AU}$, the WJ/HJ ratio is $\sim 1.6$.  Accounting for selection effects would further increase the WJ/HJ ratio.

The observed WJ/HJ ratio is thus in contradiction with most population synthesis results.  Considering migration due to Lidov-Kozai oscillations from a planetary companion, \cite{petrovich2016} produce roughly twice as many HJs as WJs.  This WJ/HJ ratio is the highest found in a population synthesis thus far,  but may result in part from the rather specific semi-major axes selected for both planets, chosen so that eccentricity oscillations are not frozen by general relativity at WJ distances.  The semi-major axis of the outer planet in particular may strongly affect the migration rate at WJ distances, because it helps determine the orbital distance at which eccentricity oscillations freeze to a large value \citep[see][Section 3.1]{kra2016}, after which the planet migrates inward to HJ territory quickly, and spends a negligibly small amount of time at WJ distances.

The above difficulties in forming WJs by high-eccentricity migration leads us to consider the possibility that most WJs form in-situ, by disk migration, or some combination of these two processes.  At typical WJ semi-major axes ($\sim 0.3$ AU), theoretical work shows that sufficiently massive rocky cores can accrete gas and undergo runaway accretion \citep{lee2014}, although growing the core quickly enough before the gas disperses may be challenging \citep{lee2016}.  In-situ formation of WJs was also recently argued by \cite{huang2016}, who found that close, rocky neighbors are common in observed WJ systems.  However, both in-situ formation and disk-driven migration have difficulty in explaining eccentric WJs.  Distinct populations of WJs have previously been proposed, with the eccentric WJs forming via some form of high-eccentricity migration, and the circular WJs forming by a different channel \citep{dawson2013,petrovich2016}.

This paper considers the scenario in which most WJs reach their current sub-AU orbits either by in-situ formation or disk migration, after which a subset of WJs undergo secular eccentricity oscillations driven by an exterior companion -- many such companions have been detected through radial velocity studies (see Section \ref{sec:observed}).  We examine the possibility of raising the eccentricities of WJs by secular interactions with distant planetary companions, so that the eccentricity varies between $e \simeq 0$ and a maximum value $e = \emax$.  In order for a WJ with observed eccentricity $\eobs$ to have its eccentricity raised by an external (and possibly undetected) companion, the maximum eccentricity must satisfy $\emax \geq \eobs$.  This places constraints on the properties of the planetary perturber, in terms of its mass, separation, inclination, and eccentricity.  We focus exclusively on secular perturbations, because in-situ scatterings have been shown to be ineffective in raising the eccentricities of close-in planets \citep{petrovich2014}.

The role of external companions in raising the eccentricities of WJs has been studied before.  However, most previous works \citep[e.g.][]{dong2014,dawson2014,antonini2016, petrovich2016} have focused on the situation where WJs achieve very small pericenter distances such that the orbit decays via tidal dissipation (and are thus in the process of becoming HJs).  If we do not require the WJs to attain such small pericenter distances, and instead focus on generating more modest eccentricities ($e \simeq 0.2 - 0.5$), the requirements on the external companion are less stringent.  Note that recent work has considered generating eccentric WJs in systems with three or more giant planets through relatively violent scattering events \citep{mustill2016}.  In contrast, in this paper we focus on systems of two widely-spaced planets where scattering does not occur, and we identify the necessary properties of external planets in generating modest eccentricities in WJs through secular processes.  This scenario requires that the outer planet have a non-zero eccentricity or inclination; such eccentricities/inclinations may result from either an initial scattering event with three or more giant planets, or perturbations from a tertiary stellar companion. Note that in order for a tertiary stellar companion to increase the eccentricity/inclination of an outer giant planet via secular interactions, it must be sufficiently close/massive so that the stellar companion induces pericenter precession in the outer planet that overcomes the precession induced by the WJ.

This paper is organized as follows. In Section \ref{sec:fiducial} we summarize our method and relevant analytic expressions for identifying the requirements for an external companion to increase the eccentricity of a WJ.  We first consider coplanar systems (Section \ref{sec:coplanar}), so that eccentricity oscillations (including the effect of an apsidal precession resonance) can be studied analytically.  We then consider inclined systems, for which octupole-level Lidov-Kozai oscillations may arise, requiring numerical integrations. Section \ref{sec:observed} considers the sample of observed and eccentric WJs with detected outer companions, and identifies the mutual inclinations necessary to raise the eccentricity of the WJ to the observed value.  In Section 4 we consider small neighboring planets to WJs, and their role in suppressing eccentricity oscillations.  We conclude in Section 5.   

\section{Secular Interactions of Warm Jupiters With Distant Planet Companions}
\label{sec:fiducial}
\subsection{Setup and Method}
We consider a system of two well-separated giant planets $m_1$ (the WJ) and $m_2$ (the exterior perturber), orbiting a star of mass $M_\star$.  We denote the semi-major axis and eccentricity of $m_1$ and $m_2$ as $\ain, \ein$ and $\aout, \eout$ respectively.  The planets may have a mutual inclination $I$, defined through $\cos I = \hatLin \cdot \hatLout$, where $\hatLin$ and $\hatLout$ are unit vectors along the angular momenta $\Lvec_\In$ and $\Lvec_\Out$.  The orbits are also specified by the eccentricity vectors $\evec_\In$ and $\evec_\Out$. For ease of notation, we frequently omit the subscript ``$\In$'', so that $e = e_\In$, $a = \ain$, etc. 

In general, we follow the evolution of $(\Lvec_\In,\evec_\In)$ and $(\Lvec_\Out,\evec_\Out)$ due to the mutual interaction between $m_1$ and $m_2$ up to octupole order, using the vector equations derived in \cite{liu2015} \citep[see also][]{petrovich2015} . The eccentricities of both planets may undergo periodic oscillations, with maximum eccentricity of the inner orbit denoted by $\emax$.  The eccentricity oscillations occur on a characteristic timescale $\tk$ (the quadrupole ``Kozai timescale''), given by
\be
\frac{1}{\tk} = \frac{m_2}{M_\star} \frac{a^3}{\aouteff^3} n_\In,
\label{eq:tk}
\ee
where we have introduced an ``effective'' outer semi-major axis,
\be
\aouteff \equiv \aout \sqrt{1 - \eout^2},
\ee
and where $n_\In = \sqrt{G M_\star/a^3}$ is the orbital mean motion of the inner planet.

Octupole effects are manifested by terms of order $\varepsilon_\oct$, where
\be
\varepsilon_\oct = \frac{M_\star - m_1}{M_\star + m_1} \frac{a}{\aout} \frac{\eout}{1 - \eout^2} \simeq \frac{a}{\aout} \frac{\eout}{1 - \eout^2}.
\label{eq:epsoct}
\ee

We also include the short-range-forces (SRFs) introduced by general relativity and tidal distortion\footnote{We do not consider the additional precession due to rotational distortion of either $M_\star$ or $m_1$, because they are both smaller than the GR term (dominant at low eccentricities) and the tidal term (dominant at high eccentricities).} of $m_1$.  These non-Keplerian potentials lead to pericenter precession and introduce two additional parameters in the equations of motion:
\be
\varepsilon_{\gr} \simeq 0.1 \bigg( \frac{M_\star}{\msun} \bigg)^{2} \bigg( \frac{m_2}{\mjup} \bigg)^{-1} \bigg( \frac{a}{0.3 \ {\rm AU}} \bigg)^{-4} \bigg( \frac{\aouteff}{3 \ {\rm AU}} \bigg)^{3},
\label{eq:epsGR}
\ee
and
\be
\begin{split}
\varepsilon_{\tide}  \simeq & 6.4 \times 10^{-5} \frac{k_{2}}{0.37} \bigg(\frac{R_{1}}{\rjup} \bigg)^5 \bigg( \frac{M_\star}{\msun} \bigg)^{2} \bigg( \frac{m_2}{\mjup} \bigg)^{-1} \\
 \times & \bigg( \frac{a}{0.3 \ {\rm AU}} \bigg)^{-8} \bigg( \frac{\aouteff}{3 \ {\rm AU}} \bigg)^{3},
\end{split}
\label{eq:epsTide}
\ee
where $R_1$ and $k_{2}$ are the radius and tidal Love number of $m_1$.
See \cite{kra2016} and \cite{liu2015} for further details and the secular equations of motion.  Since we focus on generating
modest eccentricities in the inner planet via secular interactions, we neglect dissipative tides, which act over much longer timescales than the timescale for eccentricity growth, and only modify the WJ orbit for pericenter distances much smaller than those of interest here.

This paper aims to explain eccentric WJs by secular perturbations from exterior giant planet companions. For an observed WJ with eccentricity $e = \eobs$, the constraint on an undetected outer companion can be obtained by calculating $\emax$ for outer perturbers with varying properties, and requiring $\emax \geq \eobs$. In a similar vein, if a WJ with $\eobs$ does have a detected companion, we can identify whether such a companion is capable of producing $\eobs$, by checking whether $\emax \geq \eobs$.  This latter idea is considered in Section 3 for observed WJs with exterior companions.

In the following we consider a  ``canonical'' WJ, with $m_1 = 1 \mjup$ and $a = 0.3$ AU, and explore various properties for the outer companion.  Sections \ref{sec:coplanar} and \ref{sec:ecc} consider coplanar systems, while Sections \ref{sec:mediuminc} and \ref{sec:highinc} consider inclined systems.  See Section \ref{sec:highinc} for the main results of Section 2.

\subsection{Coplanar Systems}
\label{sec:coplanar}
We begin with coplanar systems ($I = 0$). The maximum eccentricity of $m_1$ (to octupole order) is completely specified by energy and angular momentum conservation \citep{lee2003}, without the need for numerical integrations of the equations of motion.  The total energy per unit mass, including the octupole-order interaction potential between $m_1$ and $m_2$ and SRF effects for $m_1$ is
\be
\Phi = \Phi_{\rm Int} + \Phi_{\rm SRF},
\label{eq:Phi}
\ee
where
\be
\begin{split}
\Phi_{\rm Int} & = \Phi_{\rm Quad} + \Phi_{\rm Oct} \\
& = \frac{\Phi_0}{8} \bigg[ -2 - 3 e^2 + \frac{15}{8} e (3 e^2 + 4) \varepsilon_\oct \cos \Delta \varpi \bigg],
\label{eq:coplanar_energy}
\end{split}
\ee
and
\be
\begin{split}
\Phi_{\rm SRF} & = \Phi_{\gr} + \Phi_{\tide} \\
& = - \frac{\varepsilon_{\gr} \Phi_0}{j}  - \frac{\varepsilon_{\tide} \Phi_0}{15 j^9} \bigg(1 + 3 e^2 + \frac{3}{8} e^4 \bigg).
\label{eq:energy_srfs}
\end{split}
\ee
Note that in Eqs.~(\ref{eq:coplanar_energy}) and (\ref{eq:energy_srfs}), we have defined
\be
\Phi_0 = \frac{G m_2 a^2}{\aouteff^3},
\ee
as well as $j = \sqrt{1 - e^2}$, and $\Delta \varpi \equiv \varpi_\In - \varpi_\Out$ (difference in longitude of pericenter of the inner and outer orbits).
Unless the eccentricity reaches extreme values ($e \gtrsim 0.9$), the SRFs are dominated by the GR contribution, and to simplify the remainder of the analytic discussion we ignore the tidal contribution (note however that we always include it in the numerical integrations presented in this paper). 

\begin{figure}
\centering 
\includegraphics[scale=0.5]{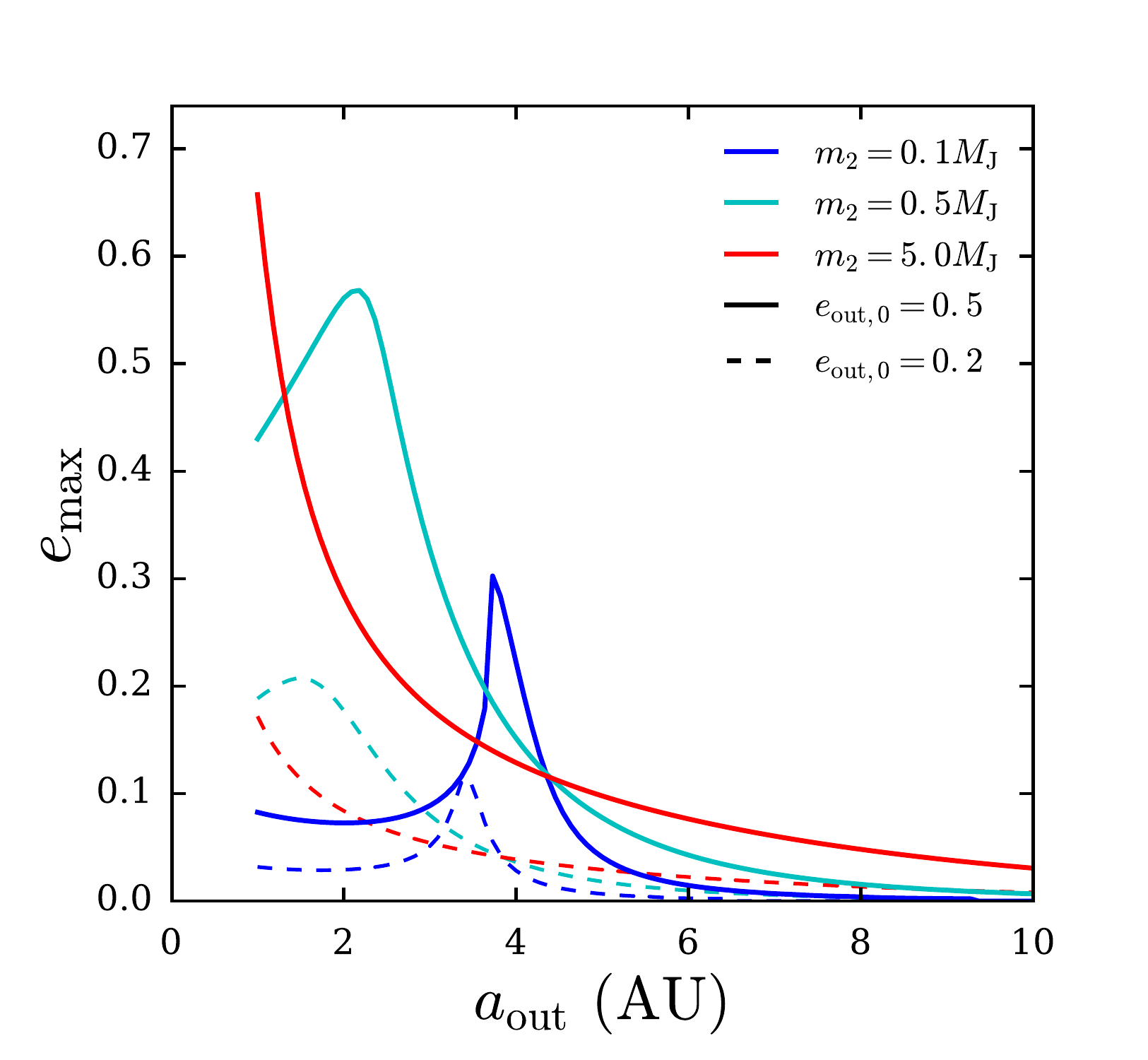}
\caption{Maximum eccentricity of the WJ as a function of $\aout$, for various  masses and eccentricities of the outer planet $m_2$.  The two planets are coplanar, with the WJ ($m_1 = 1 \mjup$) placed at $a = 0.3$ AU.  The initial eccentricity of $m_2$ is $e_{\Out,0} = 0.5$ (solid), and $e_{\Out,0} = 0.2$ (dashed). The curves for the low mass perturbers ($m_2 = 0.1, 0.5 \mjup$) have a distinctive spike, corresponding to an apsidal precession resonance, where $\dot{\varpi}_\In / \dot{\varpi}_\Out \simeq 1$.  For the $5 \mjup$ perturber, the resonance can only be achieved at small separations, where the secular approximation is no longer valid.}
\label{fig:emax_coplanar}
\end{figure}

Figure \ref{fig:emax_coplanar} shows the maximum eccentricity ($\emax$) of the canonical WJ (with $m_1 = 1 \mjup$, $a=0.3$ AU, and initial eccentricity $e_0\simeq 0$) as a function $\aout$ for various outer planet masses and eccentricities.  In general, $\emax$ increases with decreasing $\aout$, except for significant peaks at certain values of $\aout$.  These peaks arise from the ``apsidal precession resonance'' \citep{liu2015b}, which occurs when the total apsidal precession rate of the inner orbit (which consists of the precession driven by $m_2$ and the GR contribution, $\dot{\varpi}_\In = \dot{\varpi}_{12} + \dot{\varpi}_{\gr}$) matches the apsidal precession rate of the outer orbit ($\dot{\varpi}_\Out = \dot{\varpi}_{21}$).  To quadrupole order, the precession frequencies due to the secular interactions between $m_1$ and $m_2$ are 
\be
\dot{\varpi}_{12} = \frac{3}{4} \alpha^3 n_\In \frac{m_2}{M_\star} \frac{j}{(1 - \eout^2)^{3/2}}
\label{eq:dot_pomega_in}
\ee
and
\be
\dot{\varpi}_{21}  = \frac{3}{8} \alpha^2 n_\Out \frac{m_1}{M_\star} \frac{2 + 3 e^2}{(1 - \eout^2)^{2}}, 
\ee
where $\alpha \equiv a / \aout$, and $n_{\In}$ ($n_{\Out}$) is the orbital mean motion of $m_1$ ($m_2$).  The precession of $m_1$ due to GR is
\be
\dot{\varpi}_{\gr} = \frac{3 n_\In}{j^2} \frac{G M_\star}{a c^2}.
\label{eq:gr}
\ee
The resonance condition $\dot{\varpi}_\In \simeq \dot{\varpi}_\Out$ yields
\be
\begin{split}
& \frac{3}{4} \alpha^3 n_\In \frac{m_2}{M_\star} \frac{j}{(1 - \eout^2)^{3/2}} + \frac{3 n_\In}{j^2} \frac{G M_\star}{a c^2} \\ 
& \simeq \frac{3}{8} \alpha^2 n_\Out \frac{m_1}{M_\star} \frac{2 + 3 e^2}{(1 - \eout^2)^2}.
\label{eq:rescondition}
\end{split}
\ee  
This resonance is only precisely defined in the limit $\ein,\eout \ll 1$, for which Eq.~(\ref{eq:rescondition}) reduces to
\be
\frac{m_2}{M_\star} \simeq \alpha^{1/2} \frac{m_1}{M_\star} - 4 \alpha^{-3} \frac{G M_\star}{a c^2}.
\label{eq:res_low_e}
\ee
In this limit, the peak eccentricity of $m_1$ is [see Eq.~(33) of \citep{liu2015b}]
\be
e_{\rm peak} = e_{\Out,0} \bigg( \frac{m_2}{m_1} \bigg)^{1/2} \alpha^{-1/4}.
\label{eq:epeak}
\ee

For moderate values of $e$ and $\eout$, the resonance becomes ``fuzzy''
because of the variations of $e$ and $\eout$ during the secular evolution.  Nevertheless, the condition $\dot{\varpi}_{\In} \simeq \dot{\varpi}_{\Out}$, with $\eout \simeq e_{\Out,0}$ and $e \sim 0$ provides a good indicator for the resonance, as long as the eccentricity of the WJ remains moderate ($\emax \lesssim 0.5$).

For increasingly massive perturbers, the resonance cannot be achieved, unless the perturber semi-major axis is small.  For the $5 \mjup$ perturber in Fig.~\ref{fig:emax_coplanar}, the resonance can only occur when $\aout$ is comparable to $a$, where non-secular effects clearly will be emerge and the stability of the system compromised.

\begin{figure}
\centering 
\includegraphics[scale=0.5]{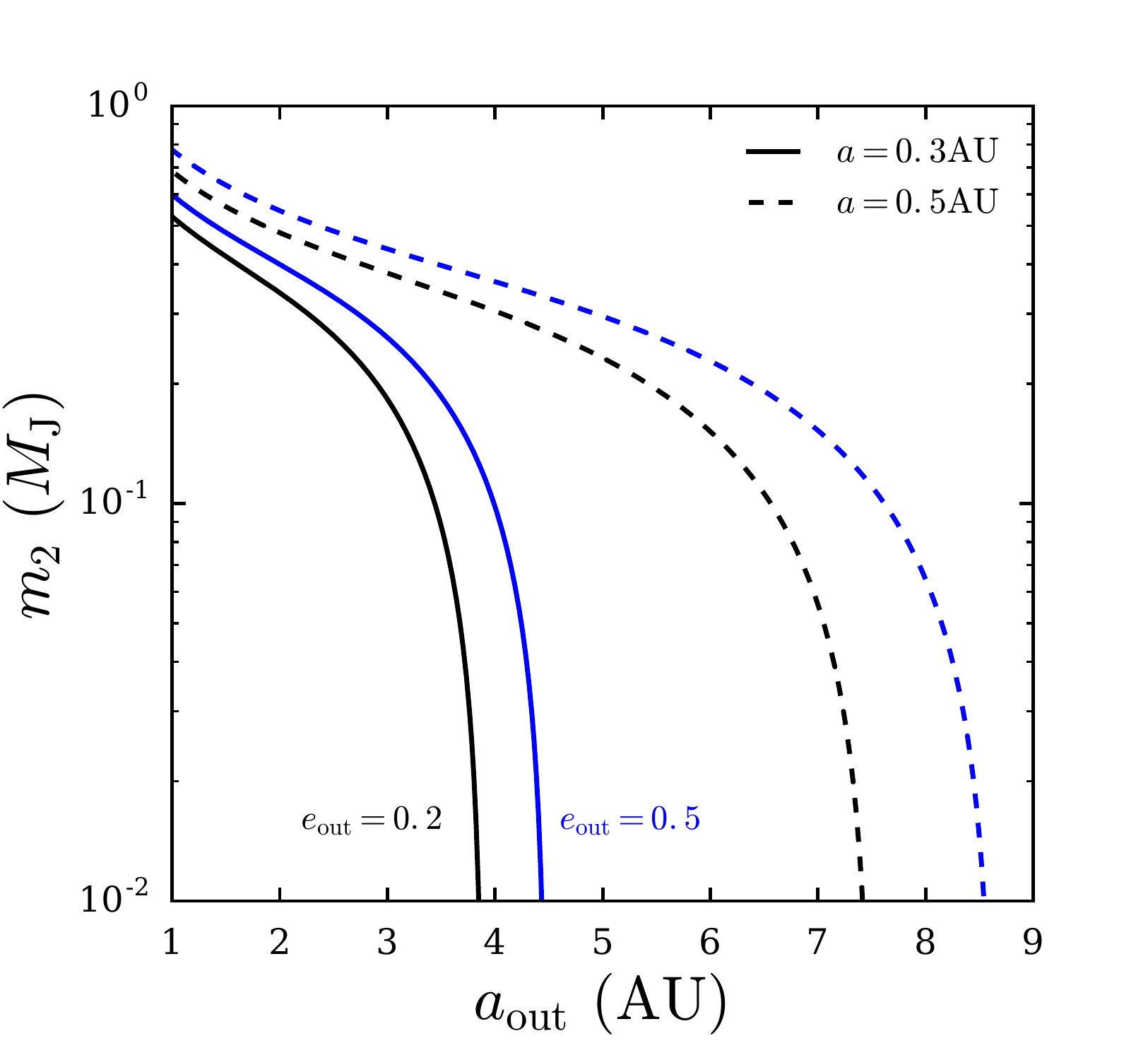}
\caption{Contours of $\dot{\varpi}_\In / \dot{\varpi}_\Out = 1$, indicating the combinations of outer planet mass $m_2$ and separation $\aout$ that may lead to an apsidal precession resonance and increased $\emax$.  In evaluating $\dot{\varpi}_\In = \dot{\varpi}_{12} + \dot{\varpi}_{\gr} $ and $\dot{\varpi}_\Out = \dot{\varpi}_{21}$ (see Eqs.~[\ref{eq:dot_pomega_in}] - [\ref{eq:gr}]), we have set $e_\In = 0$ and $e_\Out = e_{\Out,0} = 0.2$ (black), and $0.5$ (blue).  The WJ has mass $m_1 = 1 \mjup$ and semi-major axis $a = 0.3$ AU (solid curves), and $a = 0.5$ AU (dashed curves).}
\label{fig:aps_res_params}
\end{figure}

To illustrate what kind of outer planet may be capable of increasing the eccentricity of the WJ through the apsidal precession resonance, Fig.~\ref{fig:aps_res_params} shows the approximate ``resonance'' condition (curves of $\dot{\varpi}_\In / \dot{\varpi}_\Out = 1$, evaluated at $e = 0, \eout = 0.2, 0.5$. Combinations of ($m_2,\aout$) close to the lines may result in eccentricity increases in the inner orbit.  However, note that the resonance does not guarantee
large $\emax$: If $e_{\Out,0}$ is too small, $e_{\rm peak}$ will necessarily be small (see Eq.~[\ref{eq:epeak}]).

\subsection{Coplanar Systems With Modest Initial Eccentricity}
\label{sec:ecc}
Here we examine how eccentricity growth in coplanar systems depend on the initial eccentricity of the inner orbit.  \cite{li2014} have previously shown that the inner planet can achieve extreme eccentricity without SRFs (the so-called
``coplanar-Kozai mechanism''). To obtain a simple criterion for large eccentricity excitation, we approximate the outer planet eccentricity as constant.  This is justified since the change in $j_\Out = \sqrt{1 - \eout^2}$ is related to the change in $j_\In = \sqrt{1 - e^2}$ through
\be
\Delta j_\Out = - \frac{m_1}{m_2} \alpha^{1/2} \Delta j_\In,
\label{eq:delta_j}
\ee
and thus, the change in $\eout$ is often small compared to the change in $e$.

Suppose the inner planet starts with an initial $e_0$ and $\Delta \varpi_0$, and attains the maximum eccentricity $\emax$ at $\Delta \varpi = 0$.\footnote{By applying ${\rm d} e/{\rm d} \Delta\varpi=0$ in the energy conservation equation, it is easy to see that the eccentricity extremum occurs at $\Delta\varpi= 0$ or $\pi$.}  Energy conservation ($\Phi={\rm constant}$; see Eq.~[\ref{eq:Phi}]) gives
\be
\varepsilon_\oct = \frac{8}{15} \bigg[\frac{3(\emax^2 - e_0^2) - 8 \varepsilon_{\gr} (j_0^{-1} - j_\Max^{-1})}{\emax(3 \emax^2 + 4) - e_0(3 e_0^2 + 4) \cos \Delta \varpi_0 } \bigg],
\label{eq:coplanar_condition}
\ee
where $j_\Max \equiv \sqrt{1 - \emax^2}$ (note that $j_\Max$ corresponds to the minimum value of $j$).  Therefore, to attain a certain value of $\emax$, we require $\varepsilon_{\oct} \geq \varepsilon_{\oct, \rm min}$, with
\be
\varepsilon_{\oct,\rm min} = \frac{8}{15} \bigg[\frac{3(\emax^2 - e_0^2) - 8 \varepsilon_{\gr} (j_0^{-1} - j_\Max^{-1})}{\emax(3 \emax^2 + 4) + e_0(3 e_0^2 + 4)} \bigg].
\label{eq:min_oct}
\ee

\begin{figure}
\centering 
\includegraphics[scale=0.5]{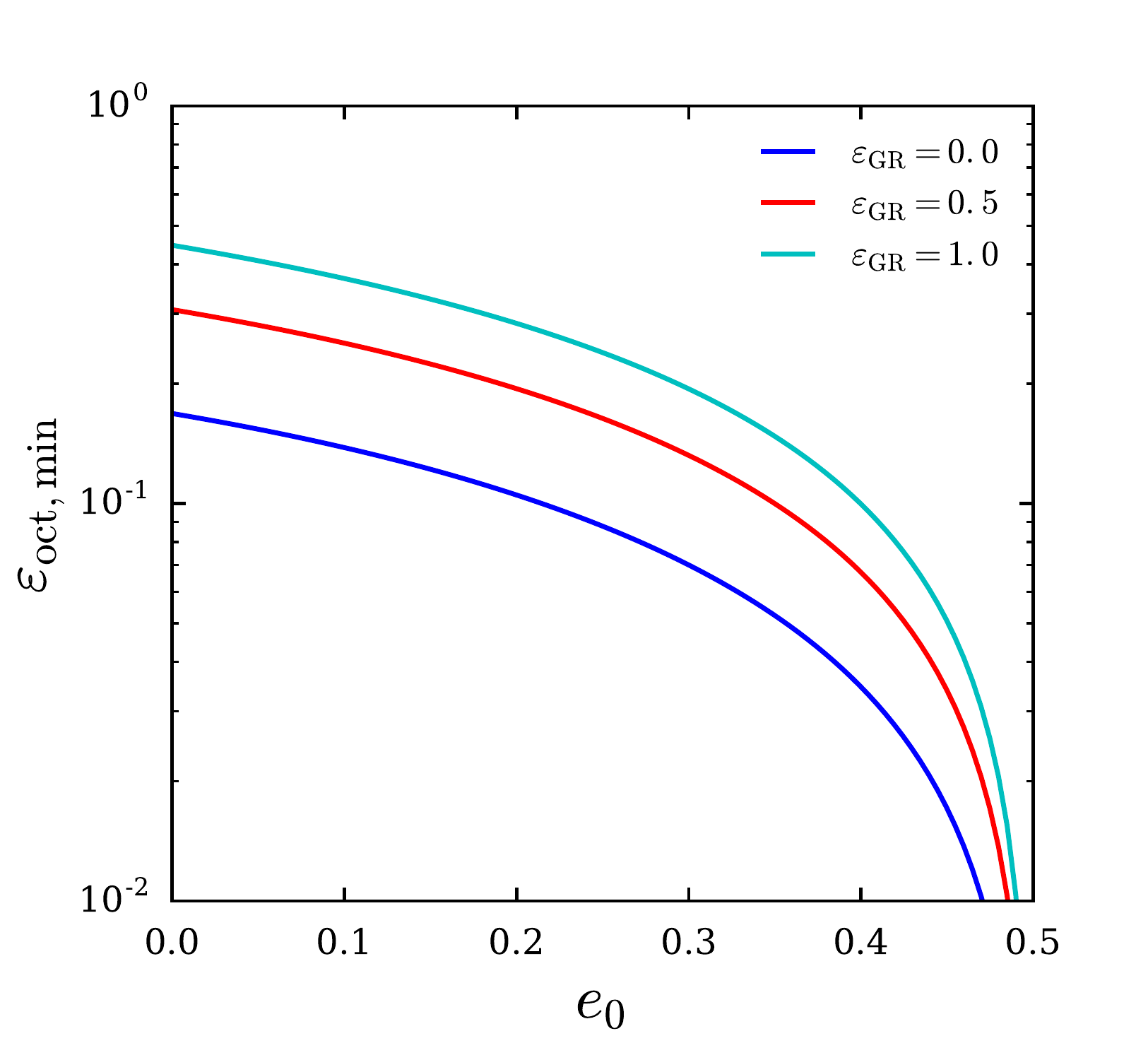}
\caption{Minimum value of $\varepsilon_\oct$ (see Eq.~[\ref{eq:epsoct}]) required to raise the eccentricity of the WJ from $e_0$ to $\emax = 0.5$; see Eq.~(\ref{eq:min_oct}).}
\label{fig:min_eps_oct}
\end{figure}

Figure 3 shows $\varepsilon_{\rm oct, min}$ as a function of $e_0$ for several values of $\varepsilon_{\gr}$.  Since this paper considers the scenario where WJs form either in-situ or by disk migration, we expect low initial eccentricities, with $e_0 \lesssim 0.1$.  As a result, we see from Fig.~\ref{fig:min_eps_oct} that the outer planet must have a strong octupole (with $\varepsilon_\oct \gtrsim 0.1$) to produce a moderate eccentricity ($\emax \sim 0.5$) in the WJ.

We briefly comment on the possibility of extreme eccentricity excitation (and the associated orbit flip) first discussed in \cite{li2014}.
To achieve $\emax \to 1$, Eq.~(\ref{eq:min_oct}) implies
\be
\varepsilon_\oct \geq \frac{8}{15} \bigg[\frac{3(1 - e_0^2) + 8 \varepsilon_{\gr} ( j_\Max^{-1} - j_0^{-1}}{7 + e_0(3 e_0^2 + 4) } \bigg].
\label{eq:flip_cond_gr}
\ee
Setting $\varepsilon_{\gr}=0$ recovers the flip condition in \cite{li2014} [see their Eq.~(14)].  Since extreme eccentricities imply $j_{\Max} \ll 1$, the large value of $\varepsilon_{\oct} $ required by Eq.~(\ref{eq:flip_cond_gr}) cannot be achieved by most dynamically stable systems.  For example, the dynamical stability condition of \cite{petrovich2015c} is
\be
\frac{\aout (1 - \eout)}{a (1 + e)} \gtrsim 2.4 \bigg[ {\rm max} \bigg (\frac{m_1}{M_\star}, \frac{m_2}{M_\star} \bigg) \bigg]^{1/3} \bigg(\frac{\aout}{a} \bigg)^{1/2} + 1.15.
\label{eq:stability}
\ee
Considering a system with $e_0 = e_{\Out,0}  = 0.5$ and $m_1/M_\star = m_2/M_\star = 10^{-3}$, and using $\varepsilon_{\gr} \simeq 10^{-2}$ and $j_{\max} = 0.1$ (note that these values lead to an extremely conservative estimate of the ratio $\varepsilon_{\gr}/j_{\Max}$), Eq.~(\ref{eq:flip_cond_gr}) implies $\aout/a \lesssim 1.2$ AU, whereas stability [Eq.~(\ref{eq:stability})] requires $\aout / a \gtrsim 5$ AU. We conclude that SRFs make extreme eccentricity excitation and orbit flipping highly unlikely for realistic systems.

\subsection{Moderately Inclined Companions}
\label{sec:mediuminc}
Next we allow the outer companion to be inclined.  When $I_0 \neq 0$, $\emax$ must be determined numerically.  The remaining results in this paper are obtained by integrating the octupole-level vector equations of motion, evolving the eccentricity and angular momentum vectors of both $m_1$ and $m_2$ \citep[e.g.][]{liu2015}. For the inner orbit we also include apsidal precession introduced by GR and tidal distortion of $m_1$.

In order to capture the octupole-order effects, the equations of motion must be integrated sufficiently long.  In all of our calculations we integrate for a timespan $10 \tk/\varepsilon_{\oct}$ (multiple ``octupole timescales'') and record the maximum value of $e$.  If the inner planet achieves a pericenter distance $a(1 - e) < \rtide \lesssim 2.7 \rjup (M_\star / m_1)^{1/3}$ \citep[e.g.][]{guillochon2011}, we terminate the integration and consider the planet tidally disrupted. 

\begin{figure}
\centering 
\includegraphics[scale=0.7]{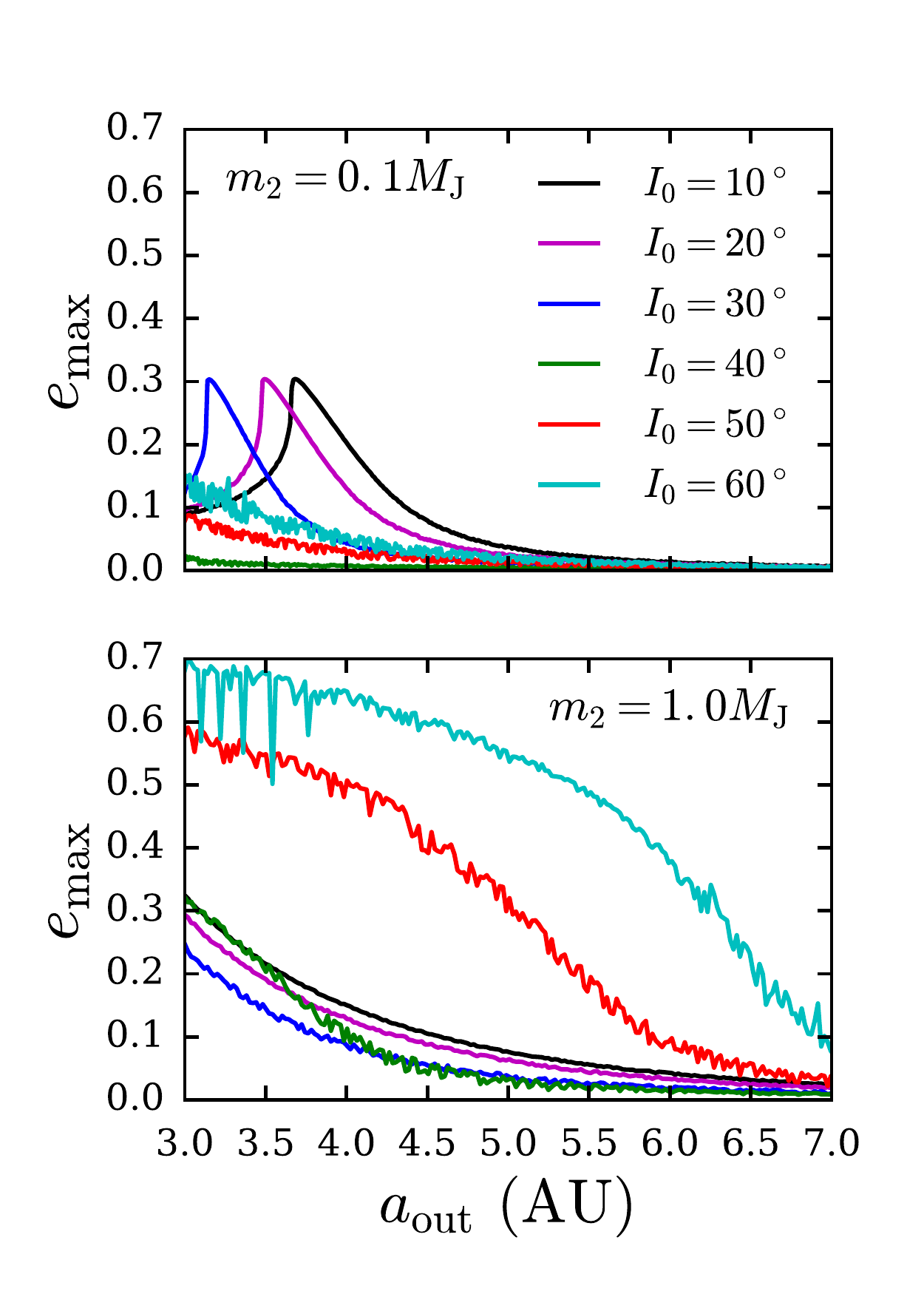}
\caption{$\emax$ versus $\aout$ for various initial inclinations, as labeled, obtained by integrating the octupole equations of motion.  We have set $e_0 = 10^{-3}$, $e_{\Out,0} = 0.5$, $a = 0.3$ AU, $m_1 = 1 \mjup$, $m_2 = 0.1 \mjup$ (top panel), and $m_2 = 1 \mjup$ (bottom panel).  Each case is initialized with $\omega_\In$ and $\Omega_\In$ randomly chosen in the range $[0, 2 \pi]$ (where $\omega_\In$ and $\Omega_\In$ are the argument of pericenter and longitude of ascending node of the inner orbit, with the invariant plane defined by the initial orbital plane of the companion).  For $m_2 = 0.1 \mjup$ and inclinations $I_0 \lesssim 30^\circ$, the behavior is qualitatively similar to the coplanar systems (see Section \ref{sec:coplanar}), with a peak eccentricity (maximum value of $\emax$) corresponding to values of $\aout$ satisfying $\dot{\varpi}_\In / \dot{\varpi}_\Out \simeq 1$.}
\label{fig:emax_inclined}
\end{figure}

We integrate a grid of inclined systems in the range $I_0 \simeq 10^\circ-60^\circ$, and vary the separation $\aout$ of the outer planet.  Figure \ref{fig:emax_inclined} shows our numerical result for $\emax$ versus $\aout$ for the various inclinations, where the inner planet properties have been set to the canonical WJ values ($m_1 = 1 \mjup$, $a = 0.3$ AU), and the perturber has initial eccentricity $e_{\Out,0} = 0.5$, and mass $m_2 = 0.1 \mjup$ (top panel) and $m_2 = 1 \mjup$ (bottom panel).  For the $0.1 \mjup$ perturber and modest inclinations ($I_0 \lesssim 30^\circ$), the behavior is similar to the coplanar systems discussed in Section \ref{sec:coplanar}.  The sharp peaks in $\emax$ exhibited in Fig.~\ref{fig:emax_inclined} occur when $\dot{\varpi}_\In \simeq \dot{\varpi}_\Out$ (cf.~Fig.~\ref{fig:emax_coplanar}).  For this set of parameters, the location of the peak eccentricity shifts to smaller $\aout$ with increasing $I_0$, until the inclination becomes large enough so that Lidov-Kozai oscillations begin.  This result, along with previous work \citep{liu2015b} shows that the apsidal precession resonance remains effective for moderately inclined systems, with $I_0 \lesssim 30^\circ$.\footnote{Similar peaks in eccentricity were seen in previous numerical calculations by \cite{ford2000} and \citep{naoz2013}.  The physical explanation of these peaks
in terms of ``apsidal precession resonance'' was first discussed in \cite{liu2015b} in the context of merging compact binaries with tertiary companions.}  

Of course, as in the coplanar case (Section 2.2), when the external companion is too massive, the resonance peak disappears (see Fig.~\ref{fig:emax_coplanar} and Eq.~[\ref{eq:res_low_e}]).

\subsection{General Inclinations: Lidov-Kozai Cycles}
\label{sec:highinc}
If the initial inclination $I_0$ is sufficiently high, Lidov-Kozai (LK) eccentricity/inclination oscillations may be induced \citep{lidov1962,kozai1962}, even when the interaction between $m_1$ and $m_2$ is truncated to the quadrupole order. (By contrast, eccentricity excitation in coplanar or low-inclination systems operates only when the octupole effect is included.) 

To quadrupole order, LK oscillations of general hierarchical triple systems, including SRFs, can be determined analytically \citep{fabrycky2007, liu2015, kra2017}.  The behavior of $\emax$ as a function of $I_0$ depends on two dimensionless parameters: the angular momentum ratio of the inner and outer orbits,
\be
\eta = \bigg( \frac{L_\In}{L_\Out} \bigg)_{e = 0} = \frac{m_1}{m_2} \sqrt{\frac{a}{\aout (1 - \eout^2)}},
\ee
and the strength of the SRFs (e.g. $\varepsilon_{\gr}$, $\varepsilon_{\tide}$; see Eqs.~[\ref{eq:epsGR}] and [\ref{eq:epsTide}]). The most general expression for $\emax(I_0)$ can be found in \citealt{kra2017} (see their Eqs.~[20] and [23]).  In particular, eccentricity excitation of the inner planet occurs when $I_0$ lies in the ``LK window'', given by
\be
(\cos I_0)_- \leq \cos I_0 \leq (\cos I_0)_+, 
\label{eq:cos_range}
\ee
where
\be
(\cos I_0 )_{+} = \frac{1}{10} \bigg(-\eta + \sqrt{60 + \eta^2- \frac{80}{3} \varepsilon_{\gr} } \bigg),
\label{eq:I_plus}
\ee
and
\be
(\cos I_0 )_{-} = 
\begin{cases}
\frac{1}{10} \bigg(\! \! \! -\! \eta - \! \sqrt{60 + \eta^2- \frac{80}{3} \varepsilon_{\gr}} \! \! \bigg), &  \! \text{if } \eta \leq 2 (1 + \frac{2}{3} \varepsilon_{\gr} ) \\    
-\frac{2}{\eta} \bigg(1 + \frac{2}{3}\varepsilon_{\gr} \bigg), & \text{otherwise}. 
\label{eq:I_min}
\end{cases}
\ee
In the above expressions, we have included only the SRF associated with GR.

When the octupole effect is included, the properties of the eccentricity-inclination oscillations cannot be determined analytically, and the relation $\emax(I_0)$ and the associated ``LK window'' can be significantly modified. Nevertheless, one analytical quadrupole result survives: The ``limiting eccentricity'' $e_{\rm lim}$, which is the peak of the $\emax(I_0)$ relation, remains valid even when the octupole terms are included \citep{liu2015,kra2017}. This $e_{\rm lim}$ (assuming $e_0 = 0$) is given by 
\be
\frac{3}{8} (j_{\rm lim}^2 - 1) \bigg[-3 + \frac{\eta^2}{4} \bigg(\frac{4 j_{\rm lim}^2}{5} - 1 \bigg) \bigg] + \bigg(\frac{\Phi_{\rm SRF}}{\Phi_0} \bigg)_{e = 0}^{e = e_{\rm lim}} = 0,
\ee
where $j_{\rm lim} = \sqrt{1 - e_{\rm lim}^2}$, and occurs at the inclination $I_{0,\rm lim}$, given by 
\be
\cos I_{0,\rm lim} = \frac{\eta}{2} \bigg(\frac{4}{5} j_{\rm lim}^2 - 1 \bigg).
\label{eq:cosIlim}
\ee
\citep[see][]{kra2017}.  Note that $e_{\rm lim}$ is not achievable if Eq.~(\ref{eq:cosIlim}) yields unphysical values of $\cos I_{0,\rm lim}$.

\begin{figure*}
\centering 
\includegraphics[width=\textwidth]{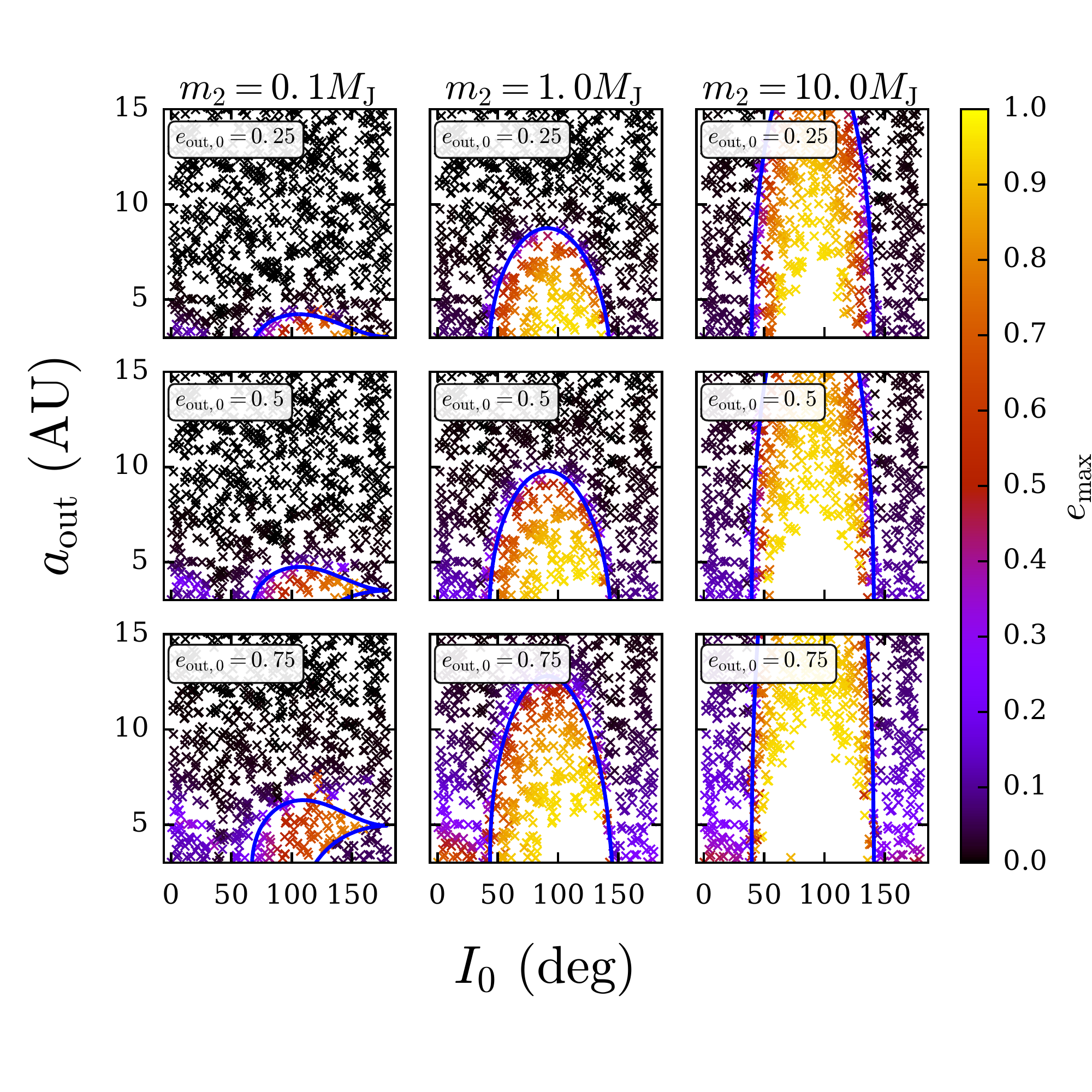}
\caption{Maximum eccentricity $\emax$, in terms of $(I_0,\aout)$ parameter space, for various outer planet masses and eccentricities, as labeled.  Each point represents a system that survives tidal disruption of the WJ (which occurs when $\emax$ is too large).  The maximum eccentricity is obtained by integrating the octupole equations of motion for a number of octupole timescales, and recording the maximum value of $e$ achieved.  The blue curves depict the quadrupole ``LK window'' for eccentricity excitation (see Eqs.[\ref{eq:cos_range}] - [\ref{eq:I_min}]).  The quadrupole prediction for the LK window is reasonably accurate for $e_{\Out,0} = 0.25, 0.5$, but fails for $e_{\Out,0}= 0.75$. }
\label{fig:emax_LK}
\end{figure*}

To examine how the quadrupole ``LK window'' (Eqs.~[\ref{eq:I_plus}] - [\ref{eq:I_min}]) may be modified by octupole, we conduct a large set of numerical integrations for the canonical WJ ($m_1 = 1 \mjup$, $a = 0.3$ AU), for perturber masses $m_2 = 0.1, 1, 10 \mjup$.  For each perturber mass, we explore several values of the initial eccentricity $e_{\Out,0}$, and sample over the full range of initial inclinations $I_0$ and a wide range of separations $\aout$.  Figure \ref{fig:emax_LK} shows the results in the ($I_0,\aout$) parameter space, where we plot the value of $\emax$ achieved over the integration span ($10 \tk/\varepsilon_\oct$).  For reference, the quadrupole ``LK window'' is also depicted, as calculated from Eqs.(\ref{eq:I_plus}) - (\ref{eq:I_min}).  Non-zero $\emax$ outside these inclination limits arises from octupole effects, either from the apsidal precession resonance (see Section \ref{sec:coplanar}) for low inclination systems, or from octupole-level LK oscillations.  For the lowest value of $e_{\Out,0}$ considered ($e_{\Out,0} = 0.25$), the systems are well-described by the quadrupole limit.  As $e_{\Out,0}$ increases, deviations from the quadrupole predictions begin to emerge, and non-zero $\emax$ may be generated well outside of the quadrupole LK window, especially when $e_{\Out,0} = 0.75$.  Notice that the results are approximately symmetric around $I_0 = 90^\circ$ when $m_2 = 1, 10 \mjup$, but exhibit considerable asymmetry when $m_2 = 0.1 \mjup$.  This arises because in the test-particle limit ($\eta \ll 1$) the equations of motion are symmetric around $90^\circ$, but this symmetry disappears when $\eta \sim 1$ (e.g. Liu et al. 2015).

Inspection of Fig.~\ref{fig:emax_LK} allows us to identify the types of outer planetary perturbers necessary to raise the eccentricity of a canonical WJ.  To generate $\emax \simeq 0.5$, relatively high ($I_0 \gtrsim 50^\circ$) mutual inclinations are needed.  A Jupiter-mass outer planet must be located within $\sim 10$ AU, unless it is extremely eccentric, with $\eout = 0.75$.  A sub-Jovian mass planet ($m_2 = 0.1 \mjup$) must be located within $\sim 3$ AU, most likely in a retrograde orbit.  Such a sub-Jovian mass perturber is therefore ineffective in generating many eccentric WJs, because only narrow ranges of separations and inclinations lead to substantial eccentricity increases.  In contrast, a massive ($\sim  10 \mjup$) perturber can generate high eccentricities at $\aout \sim 15$ AU and beyond.

\begin{figure*}
\centering 
\subfigure[]{
	\label{subfig:correct}
	\includegraphics[width=\textwidth]{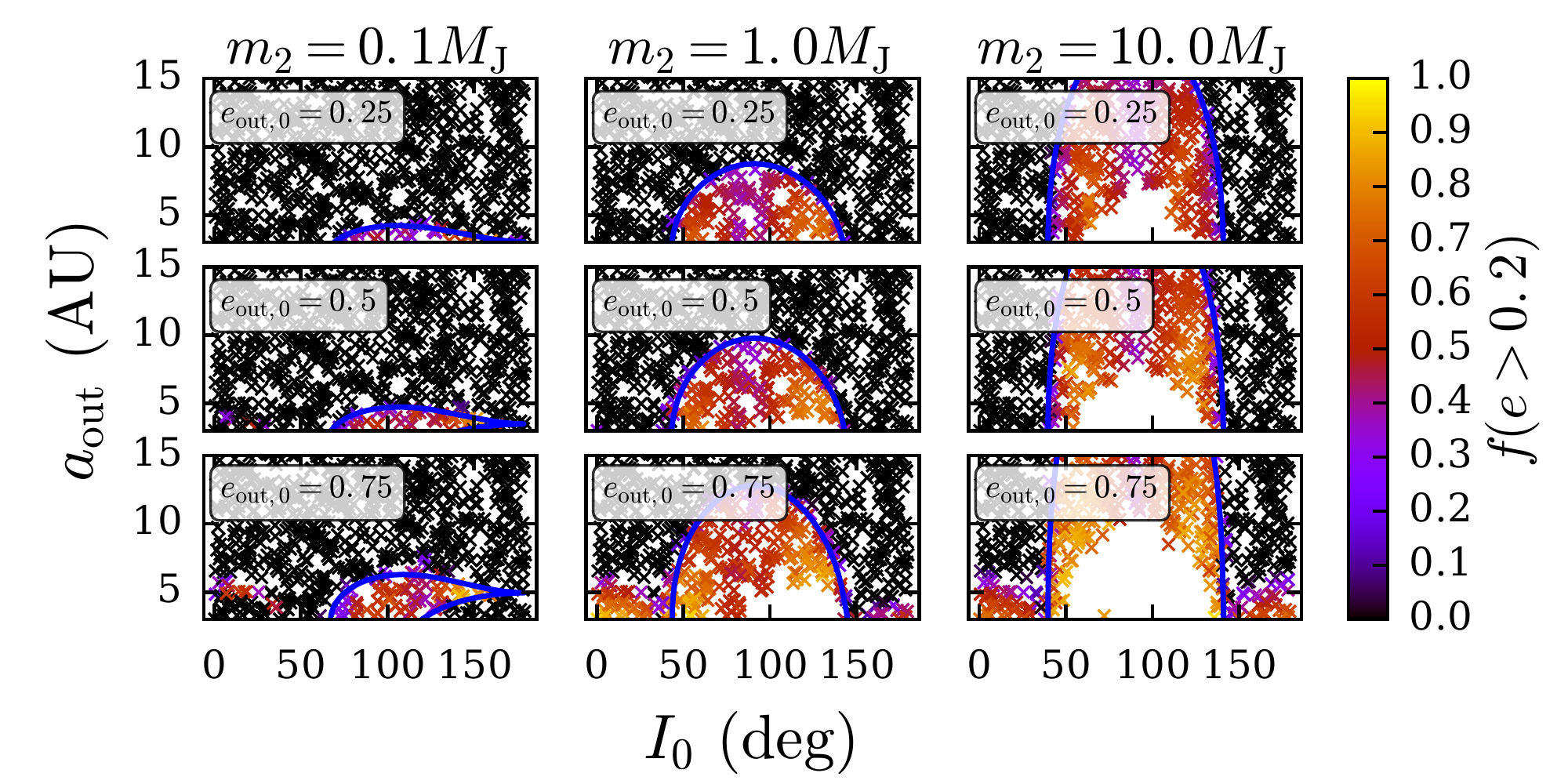} } 

\subfigure[]{
	\label{subfig:correct}
	\includegraphics[width=\textwidth]{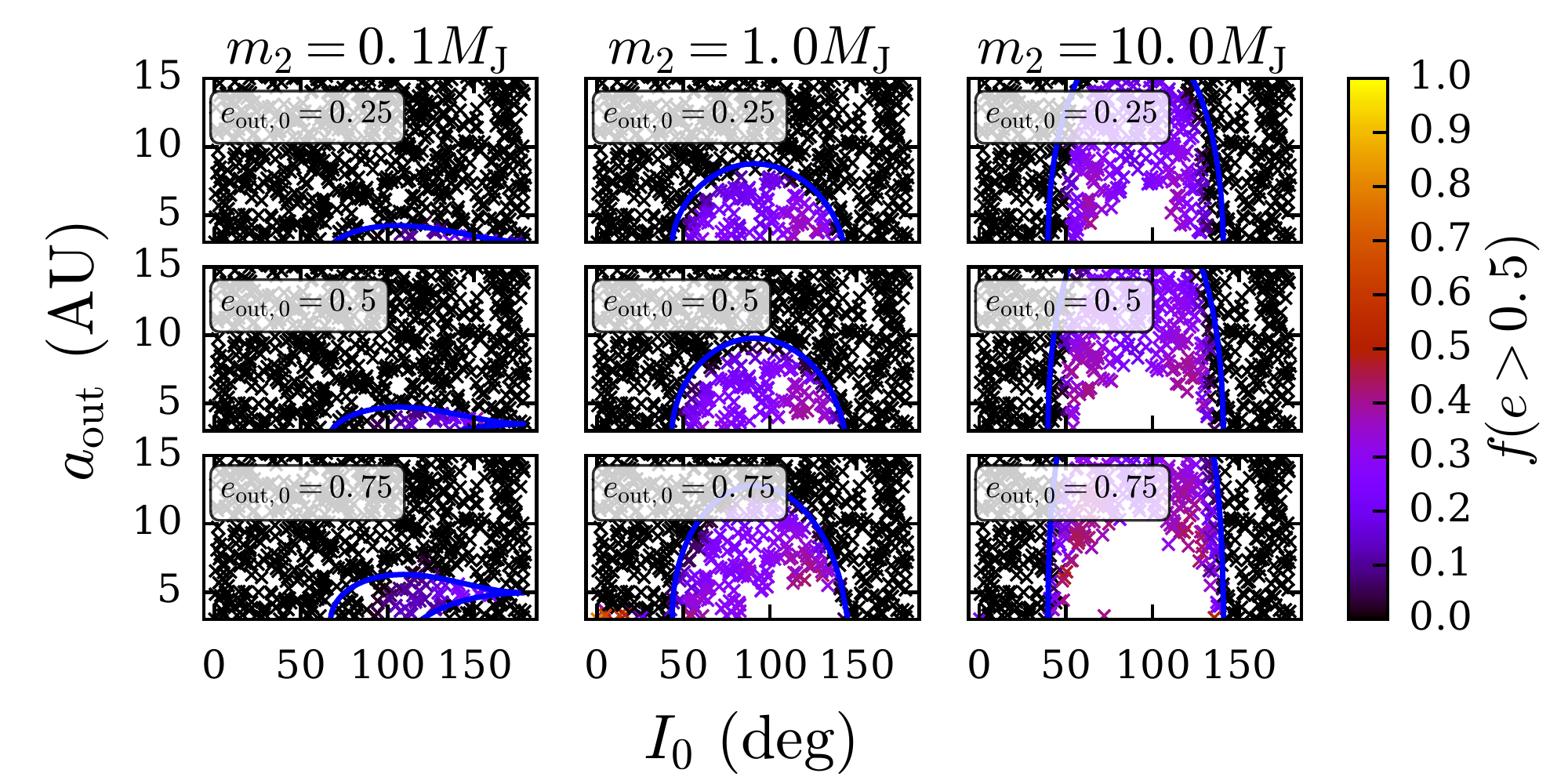} } 
\caption{(a): Same numerical experiments as depicted in Fig.~\ref{fig:emax_LK}, but showing the fraction of the total integration time that the WJ spends with $e$ above $0.2$. Note that $f(e > 0.2)$ is relatively high, often $\gtrsim 0.5$.  (b): Same as (a), but showing the fraction of time spent above $e = 0.5$.  Note that $f(e > 0.5)$ is typically less than $\sim 0.2$.}
\label{fig:frac_LK}
\end{figure*}

Figure \ref{fig:frac_LK} depicts the same numerical experiments as in Fig.~\ref{fig:emax_LK}, but shows the fraction of the total integration time that the WJ spends above a specified eccentricity.  Figure \ref{fig:frac_LK}a shows the fraction of time spent above $e = 0.2$ [$f(e > 0.2)$], and Fig.~\ref{fig:frac_LK}b shows the fraction of time spent above $e = 0.5$ [$f(e > 0.5)$].  The fraction of time spent above $e = 0.2$ is relatively high ($\gtrsim 0.5$) for many separations and inclinations, as long as the perturber mass is $1 \mjup$ or greater.  The fraction of time spent above $e = 0.5$ is much lower, usually not exceeding $\sim 0.2$.  

We conclude that external giant planet perturbers are often effective in generating mild ($\sim 0.2$) eccentricities in WJs at low mutual inclination, but in order to produce moderate ($\sim 0.5$) eccentricities in WJs requires a relatively high inclination.  Furthermore, even with a high inclination, generating a moderate eccentricity in the WJ orbit may be difficult, because of the small fraction of time the WJ spends at or above such an eccentricity.

\section{Observed WJ Systems with Exterior Companions}
\label{sec:observed}
\subsection{Sample Description and Method}
The results of Section 2 demonstrate the types of perturber necessary in generating eccentricity in a WJ with canonical properties ($m_1 = 1 \mjup$, $\ain = 0.3$ AU).  We now consider the observed radial velocity sample of WJs with giant planet companions, and evaluate the prospects for the exterior planet to raise the eccentricity of the WJ to the observed value $\eobs$.  This sample consists of 21 systems, and is given in \citealt{antonini2016} (see their Table 1).  These systems have measured minimum masses, semi-major axes, and eccentricities for both the inner and outer planets, but lack information on the mutual inclination between $m_1$ and $m_2$. 

Several of the two-planet systems in the \cite{antonini2016} sample are sufficiently non-hierarchical (with $a_\Out/a_\In < 10$) such that the (purely secular) results described in this paper may not apply. We immediately exclude systems satisfying $a_\Out/a_\In < 3$, as non-secular effects will likely dominate.  This reduces the sample from 21 to 15 systems. We conduct an additional (albeit less extensive) set of N-body integrations for the remaining systems, and look for changes in semi-major axis of either orbit (indicative of non-secular effects).  We use the N-body code REBOUND \citep{rein2012}, and include the apsidal precession from GR and tidal distortion of $m_1$ using the REBOUNDX library\footnote{https://github.com/dtamayo/reboundx}. 

In all numerical experiments we set $a$ and $a_\Out$ equal to the observed values, uniformly sample the argument of pericenter and orbital node of each planet in the range $[0, 2 \pi]$, and sample the initial inclination between $m_1$ and $m_2$ in the range $I_0 = [0,\pi]$.  We explore various possibilities for the planet masses $m_1$ and $m_2$ and initial values of $e$ and $\eout$, as described below.  The integration times are the same as described in Section \ref{sec:mediuminc}, and we record the maximum value of $e$, as well as the fraction of time the system spent with $e \geq e_{\rm obs}$ [denoted as $f(e \geq e_{\rm obs})$].  

\subsection{Fiducial Experiment}
\label{sec:obsfiducial}
\begin{figure*}
\centering 
\includegraphics[width=\textwidth]{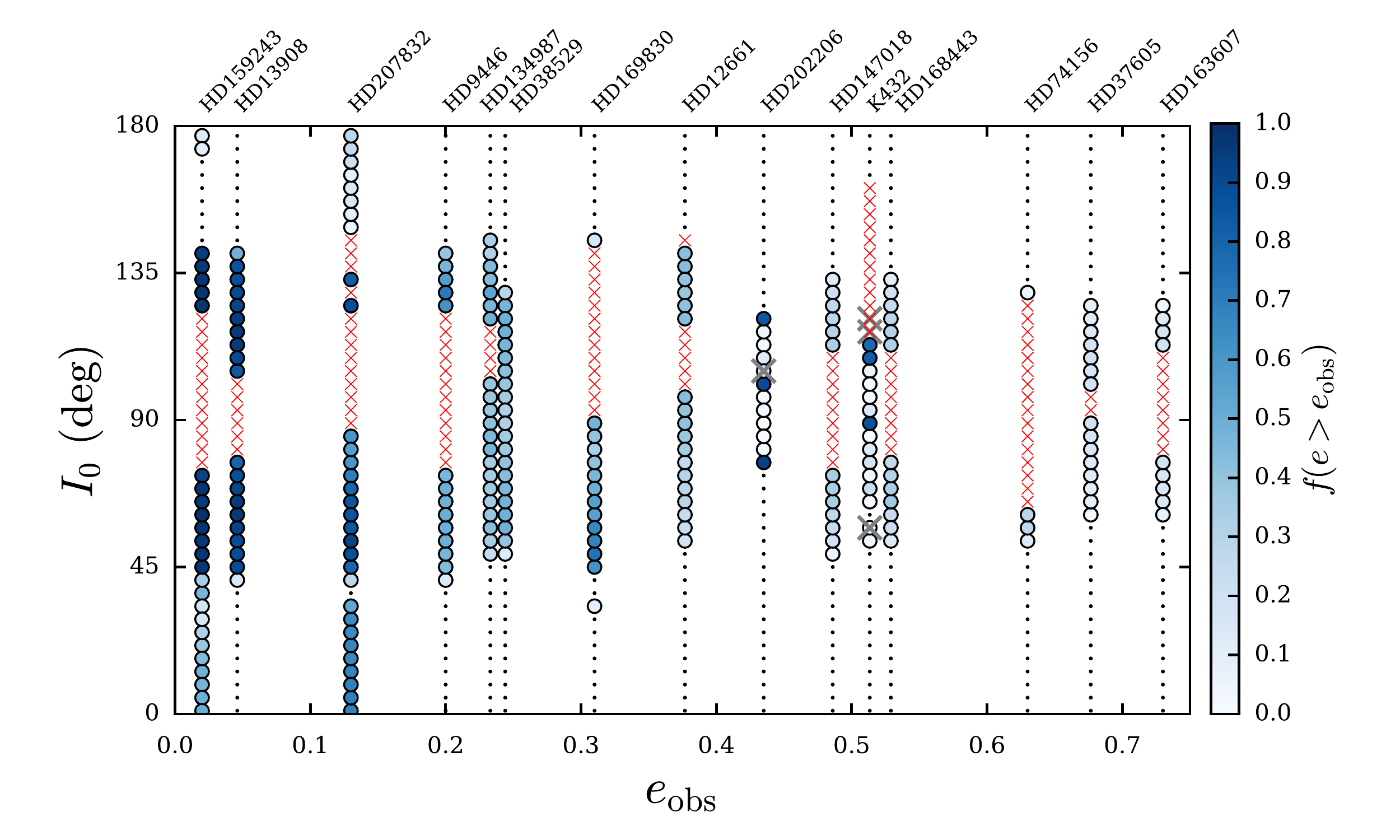}
\caption{Constraints on the required mutual orbital inclination of observed WJs with external companions (see \citealt{antonini2016}, Table 1 for the system parameters).  The results are obtained through numerical integrations, including apsidal precession from GR and tides.  If $a_\Out / a_\In < 10$ we conduct N-body integrations, while if $a_\Out / \In > 10$ we conduct secular integrations. The outcome of the integration is indicated by the symbol type.  Small red crosses: tidally disrupted (i.e. forbidden inclinations). Black dots: $\emax < \eobs$. Blue circles: $\emax > \eobs$ (i.e. the inclinations needed to generate the observed eccentricity).  The color scale indicates the fraction of time the system spent with eccentricity $e \geq e_{\rm obs}$.  The large grey crosses depicted in the results for HD202206 and Kepler-432 indicate N-body integrations where either $a_\In$ or $a_\Out$ changed by more than 10 percent, indicative of non-secular effects. In most cases, mutual inclinations $I_0 \gtrsim 40^\circ - 50^\circ$ are needed to generate the observed WJ eccentricity.  This is in agreement with the results for the canonical WJ system considered in Section \ref{sec:highinc}.}
\label{fig:wjdata_fiducial}
\end{figure*}

First we conduct a fiducial set of experiments assuming that the inner planet orbit is initially circular, while the outer planet has the initial $e_{\Out,0}$ given by the observed value, and the observed minimum masses for $m_1$ and $m_2$ are equal to the true masses. Figure \ref{fig:wjdata_fiducial} depicts results for a grid of inclinations.  We split the results into three possible outcomes:
$\emax \leq e_{\rm obs}$, $\emax \geq e_{\rm obs}$, and tidal disruption.  The color scale indicates the fraction of the total integration time spent with eccentricity exceeding the observed value [$f(e \geq \eobs)$].  In all but two systems (HD159243 and HD207832), high mutual inclinations ($I_0 \gtrsim 40^\circ - 50^\circ$) are needed to produce the observed eccentricity.

Although Fig.~\ref{fig:wjdata_fiducial} presents a qualitative picture of the necessary initial inclinations, it is incomplete because each inclination corresponds to a particular set of initial orbital phases. Thus, we present a large set (1000 trials) of numerical integrations, sampling the full range of precession phases ($\omega_\In$, $\Omega_\In$, $\omega_\Out$) and mutual inclinations.  For each observed system, the trials that led to $\emax \geq \eobs$ (without resulting in tidally disruption) are plotted in Fig.~\ref{fig:wjdata_fiducial_allphase}, showing $f(e \geq \eobs)$ versus $I_0$.  In nearly all cases with $\eobs \gtrsim 0.2$, a mutual inclination greater than about $40^\circ - 50^\circ$ is required to generate the observed eccentricity.  The exterior companions simply do not have sufficient octupole strengths for a coplanar configuration to drive eccentricity oscillations of sufficient amplitude in the WJ, and instead require high inclinations so that LK oscillations are induced.  Two exceptions are HD159243 and HD207832.  The observed eccentricities of both of these WJs are readily explained with coplanar configurations because of the relatively low values ($\eobs \simeq 0.02$ and $0.1$ respectively).
 
As discussed in Section \ref{sec:highinc}, in order for secular eccentricity oscillations from exterior companions to be a plausible explanation for eccentric WJs, we also require that the system spend a sufficiently large fraction of time with $e \geq \eobs$.  The quantity $f(e \geq \eobs)$ has a complicated dependence on inclination and system parameters, and must be examined on a case-by-case basis (see Fig.~\ref{fig:wjdata_fiducial_allphase}).  As expected, systems with higher $\eobs$ usually have lower values of $f(e \geq \eobs)$.  The two systems with the highest eccentricities (HD37605 and HD163607) have $f(e \geq \eobs) \lesssim 0.2$ for all inclinations.

We note that unlike in Fig.~\ref{fig:wjdata_fiducial}, all the results in Fig.~\ref{fig:wjdata_fiducial_allphase} were obtained by integrating the secular equations of motion, without accompanying N-body calculations for the less hierarchical systems.  These results thus may not completely capture the full physical behavior for some of the less hierarchical systems, especially Kepler-432 (the system that exhibits occasional non-secular behavior in our N-body integrations shown in Fig.~\ref{fig:wjdata_fiducial}).    However, note that Kepler-432 is a WJ orbiting an evolved star \citep{ciceri2015, ortiz2015, quinn2015}, and the large stellar radius may lead to enhanced tidal interactions and possibly dissipation in the star and orbital decay.  The results for Kepler-432 should therefore be taken with caution, since the treatment in this paper does not include these additional physical ingredients.

\subsection{Additional Numerical Experiments}
\label{sec:additional}
\begin{figure*}
\centering 
\includegraphics[width=\textwidth]{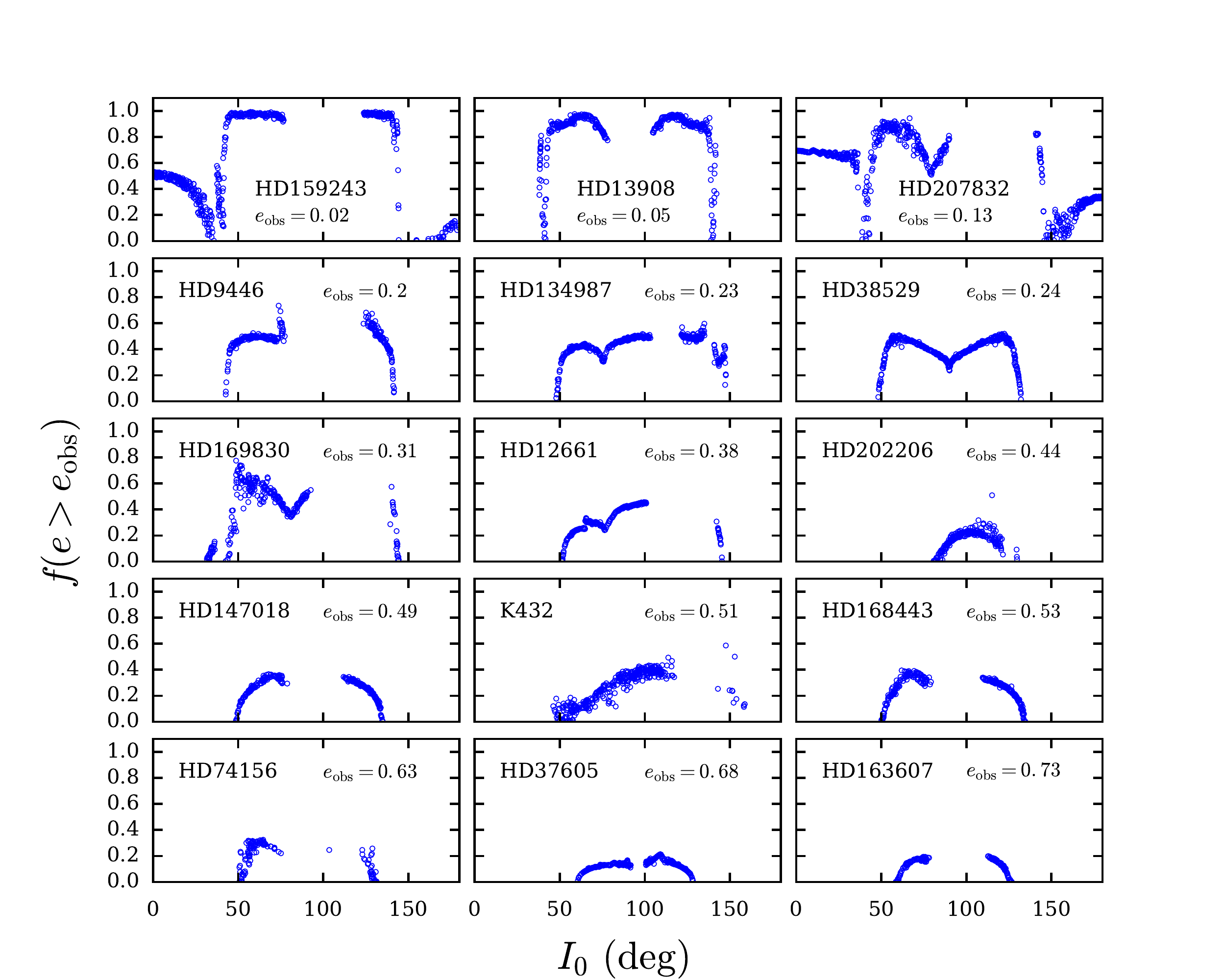}
\caption{{\it Fiducial experiment (Section \ref{sec:obsfiducial}):} Large set ($\sim 1000$ trials) of numerical integrations of observed WJ systems with external companions, with inclinations and orbital angles randomly sampled (see Table 1 for further information).  For each set of initial conditions, we integrate the secular equations of motion, and calculate the fraction of time that the WJ spends at an eccentricity greater than the observed value [$f (e > \eobs)$].  The dependence of $f$ with initial inclination varies from system-to-system, and is often complex.  High mutual inclinations are usually needed to generate the observed eccentricity, in agreement with Fig.~\ref{fig:wjdata_fiducial}. }
\label{fig:wjdata_fiducial_allphase}
\end{figure*}

Next we investigate how the results of the fiducial experiments (Section \ref{sec:obsfiducial}) may change when several of the assumptions are modified. We repeat the experiments depicted in Fig.~\ref{fig:wjdata_fiducial_allphase}, but vary the following:

\begin{itemize}
\item We allow the initial eccentricity of the WJ to be non-zero.  A WJ that formed either in-situ or underwent disk migration is expected to begin with low eccentricity, but here we allow for an initial value $e_0 = 0.1$.  Such an eccentricity may conceivably be induced by planet-disk interactions \citep[e.g.][]{goldreich2003,tsang2014,duffell2015}, or perhaps from a scattering event with another body early in the system's history.  We denote this set of experiments as {\texttt Eccentric-in} (with all other parameters identical to the fiducial set).

\item We consider the possibility that the outer planet initially had a higher eccentricity than the observed value.  If both planets are observed at a random point in a mutual eccentricity oscillation cycle, then the initial eccentricity of the outer planet may have been higher.  For a coplanar system, the change in $j_\Out = \sqrt{1 - \eout^2}$ is related to the change in $j_\In = \sqrt{1 - e_\In^2}$ via Eq.~(\ref{eq:delta_j}).
Since some of the observed systems are not exceedingly hierarchical, there may be a moderate change in $\eout$ over the eccentricity oscillation cycle.  To explore this possibility, we repeat the fiducial experiments but increase the outer eccentricity by $0.1$ relative to the observed value.  Thus, the initial outer eccentricity is $e_{\Out,0} = e_{\Out,\rm obs} + 0.1$ (keeping the other parameters identical to the fiducial set).  We denote this set of experiments as {\texttt Eccentric-out}.

\item We note that the observed masses are only lower limits.  A higher value for the outer planet mass may lead to a higher eccentricity for the inner planet.  To examine this possibility in a simple manner, we increase the outer planet mass by a factor of two: thus $m_2 = 2 (m_2 \sin{i_2})_{\rm obs}$. We denote this set of experiments as {\texttt Increase-mass-out}.

\item We increase the observed inner planet mass by a factor of two: thus $m_1 = 2 (m_1 \sin{i_1})_{\rm obs}$.   We denote this set of experiments as {\texttt Increase-mass-in}.

The parameters adopted for these experiments are summarized in Table 1.  Note that all these experiments except {\texttt Increase-mass-in} provide a more optimistic scenario in producing eccentric WJs compared to the fiducial case.
\end{itemize}

\begin{table*}
 \centering
 \begin{minipage}{180mm}
\caption{Various sets of numerical experiments involving observed WJs with outer planetary companions (see Sections \ref{sec:obsfiducial}, \ref{sec:additional}, and Fig.~\ref{fig:wjdata_all_experiments}).  The data set is given in \citealt{antonini2016} (see their Table 1). For each observed system, both planets have measured eccentricities, semi-major axes, and minimum masses.  For all experiments we set $a_\In$, $a_\Out$ to the observed values, and randomly sample the argument of pericenter and node ($\omega, \Omega$) of both planets in the range $[0- 2 \pi]$, and the mutual inclination of the planets in $I_0 = [0, \pi]$.  For each experiment we conducted 1000 numerical integrations, out which a small subset (less than $20 \%$) resulted in tidal disruption of the inner planet.}
  \begin{tabular}{@{}lllll@{}}
  \hline
  Name & $m_1$  & $m_2$ & $e_{\In,0}$ & $e_{\Out,0}$  \\
 \hline

{\texttt Fiducial} & $(m_1 \sin{i_1})_{\rm obs}$ & $(m_2 \sin{i_1})_{\rm obs}$ & 0.001 & $e_{\Out,\rm obs}$ \\
{\texttt Eccentric-in} & $(m_1 \sin{i_1})_{\rm obs}$ & $(m_2 \sin{i_1})_{\rm obs}$ & 0.1 & $e_{\Out,\rm obs}$ \\
{\texttt Eccentric-out} & $(m_1 \sin{i_1})_{\rm obs}$ & $(m_2 \sin{i_1})_{\rm obs}$ & 0.001 & $e_{\Out,\rm obs}$ + 0.1 \\
{\texttt Increase-mass-in} & $2 (m_1 \sin{i_1})_{\rm obs}$ & $(m_2 \sin{i_1})_{\rm obs}$ & 0.001 & $e_{\Out,\rm obs}$ \\
{\texttt Increase-mass-out} & $(m_1 \sin{i_1})_{\rm obs}$ & $2 (m_2 \sin{i_1})_{\rm obs}$ & 0.001 & $e_{\Out,\rm obs}$ \\

\label{popsynthtable}
\end{tabular}
\end{minipage}
\end{table*}

\begin{figure*}
\centering 
\includegraphics[width=\textwidth]{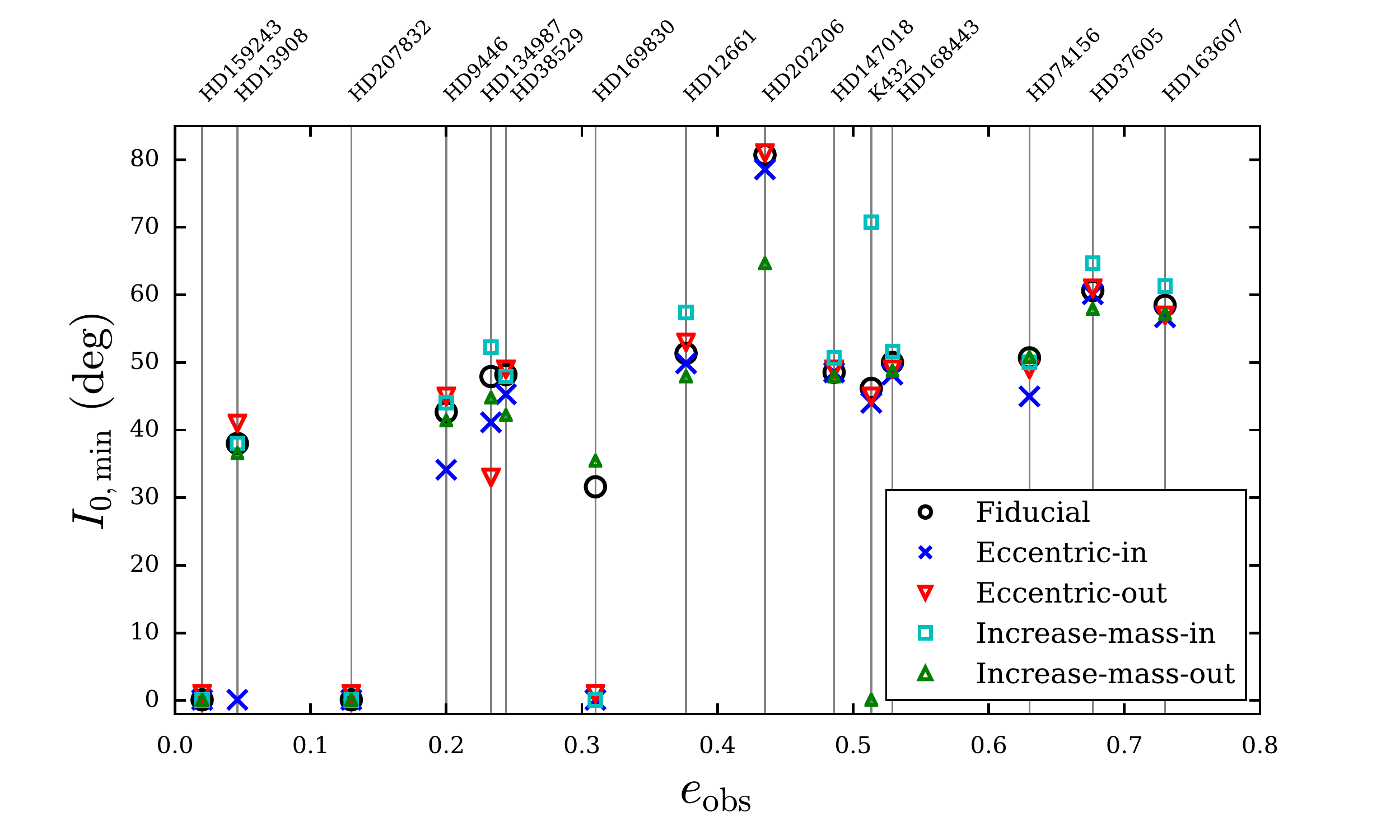}
\caption{{\it All numerical experiments: } Comparison of the various experiments (see Table 1) involving observed WJs with external companions.  Each experiment adopts different assumptions on the starting eccentricities and masses of both planets, to address the uncertainties in the initial conditions and sky-projected orientations of the orbits. For each system and experiment, we plot the minimum inclination $I_{0,\rm min}$ that led to $\emax > \eobs$, determined from integrating 1000 systems (with initial precession angles and mutual inclinations sampled randomly).  For most systems, $I_{0, \rm min}$ is not strongly affected by the experiment assumptions.  See the text for further discussion. }
\label{fig:wjdata_all_experiments}
\end{figure*}

For each experiment, we proceed exactly as in the fiducial experiment (Section \ref{sec:obsfiducial}), generating 1000 initial conditions with initial inclinations and precession phases randomly sampled over the full ranges.  In the interest of space, we omit figures analogous to Fig.~\ref{fig:wjdata_fiducial_allphase}, and instead show the minimum initial inclination needed to generate $\emax \geq \eobs$ (denoted as $I_{0,\rm min}$) in Fig.~\ref{fig:wjdata_all_experiments}.  

In most cases, $I_{0, \rm min}$ does not differ substantially from the fiducial case: inclinations greater than about $40^\circ$ are usually needed to achieve $\emax \geq \eobs$.  Two exceptions are HD169830 and Kepler-432: although the fiducial experiments imply minimum inclinations of $\sim 30^\circ$ and $\sim 50^\circ$, these additional experiments show that coplanar configurations may lead to the observed WJ eccentricity (however, see the discussion at the end of Section \ref{sec:obsfiducial} about Kepler-432.).

In summary, we find that in order for the eccentricities of the observed WJs with external companions to have arisen from secular perturbations from the outer planet, the two planets must have a mutual inclination of at least $40^\circ - 50^\circ$ in most cases.  This result is robust across various numerical experiments involving different assumptions on the initial eccentricities and masses of both planets.  The exceptions are HD159243, HD207832, and (depending on the assumptions for the initial eccentricities and masses) possibly HD169830 and Kepler-432 -- these systems can be explained with coplanar or low inclination configurations.  There is a difficulty in explaining the most eccentric WJs in the sample, because the fraction of time spent at or above $\eobs$ is low (less than $\sim 0.2$).

\section{Suppression of Eccentricity Oscillations by Close Rocky Neighbors}
Here we consider WJs with close, rocky ``neighbors,'' in addition to a distant external giant planet.   \cite{huang2016} recently found that $\sim 50 \%$ of WJs have nearby low-mass neighbors; such neighbors may lead to orbital precession of the WJ that is faster than that due to the distant giant planet, thereby suppressing eccentricity growth.

We denote the neighboring planet mass as $m'$, and the WJ and external giant planet companion haves masses $m_1$ and $m_2$, as before \footnote{We will refer to $m'$ as the ``neighbor'' and $m_2$ as the ``perturber.''}.  The planet $m'$ has semi-major axis $a'$, and may orbit interior or exterior to $m_1$, but is always interior to $m_2$.  For simplicity, we assume that $m'$ is circular and coplanar with $m_1$.  This yields a rouch estimate on the ability of $m'$ to suppress eccentricity oscillations in $m_1$.  Identifying the precise influence of $m'$ on the eccentricity of $m_1$ requires N-body integrations of three-planet systems and is beyond the scope of this paper.

In order for $m_2$ to raise the eccentricity of $m_1$, the orbital precession of $m_1$ due to $m'$ (denoted here as $\dot{\omega}$) must be smaller than the orbital precession of $m_1$ due to $m_2$ (of order $\tk^{-1}$).  We thus require
\be
\epsilon \equiv \frac{\dot{\omega}}{\tk^{-1}} \lesssim 1,
\ee
with $\epsilon$ given by
\be
\epsilon = 
\begin{cases}
\frac{m'}{m_2}  \frac{\aouteff^3}{a'^2 a}  b_{3/2}^{(1)} ( a/a') , &  \! \text{if } a' > a \\    
\frac{m'}{m_2}  \frac{a' \aouteff^3}{a^4} b_{3/2}^{(1)} (a'/a) , & \! \text{if } a' < a,
\label{eq:neighbor}
\end{cases}
\ee
where $b_{3/2}^{(1)}(\alpha)$ is a Laplace coefficient.  As a result, for specified properties of a WJ and giant planet perturber, there is a maximum value of $m'$ allowing eccentricity oscillations of $m_1$ ($m_{\rm crit}'$, obtained from setting $\epsilon = 1$).

\begin{figure*}
\centering 
\includegraphics[width=\textwidth]{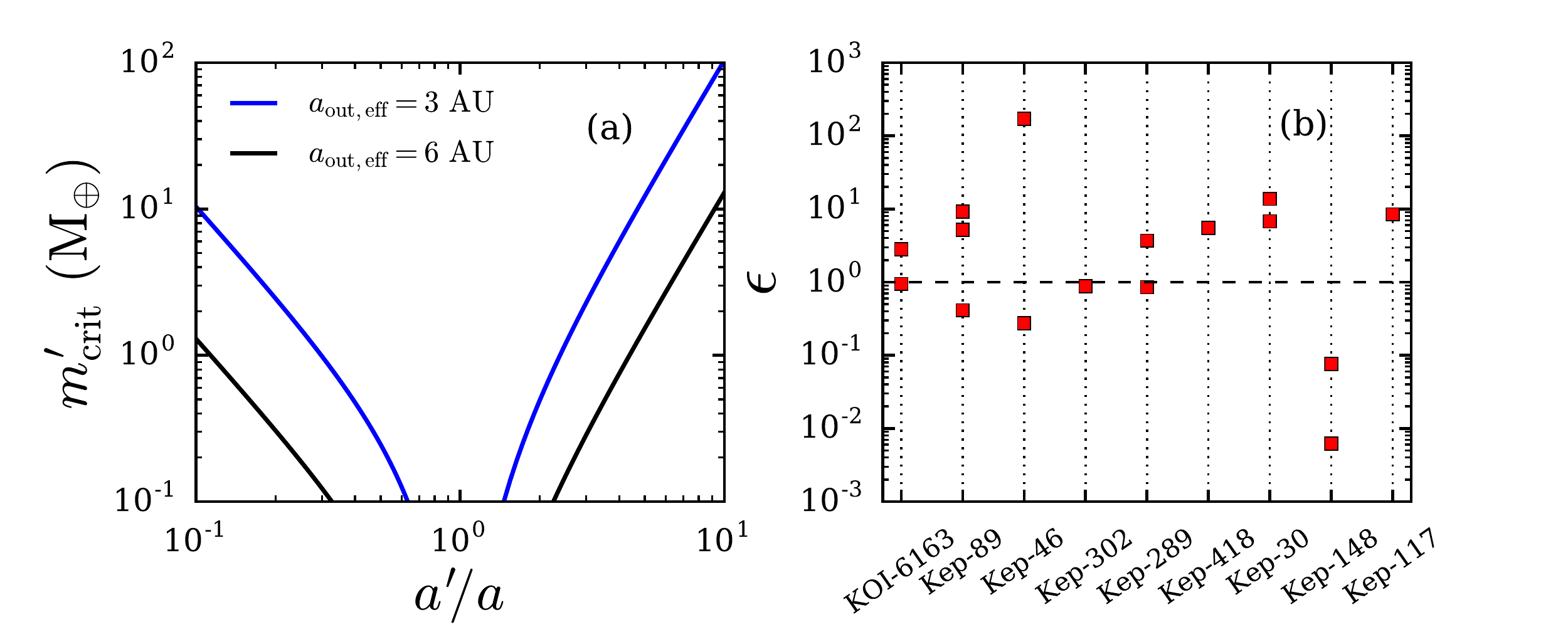}
\caption{(a): Maximum mass of $m'$ that allows eccentricity oscillations of $m_1$ (due to $m_2$), as a function of $a'/a$.  $m_{\rm crit}'$ is determined by setting $\epsilon = 1$; see Eq.~(\ref{eq:neighbor}).   The WJ has $m_1 = \mjup$, $a = 0.3$ AU, and the perturber has $m_2 = \mjup$, $\aouteff = 3, 6$ AU, as labeled. (b):  $\epsilon$ for the sample of WJs with close companions, from Huang et al. (2016).  We have set the mass and separation of a giant planet perturber to $m_2 = \mjup$ and $\aouteff = 5 a$, where $a$ is the measured WJ semi-major axis.  Nearly all systems have at least one neighboring planet that satisfies $\epsilon \gtrsim 1$, indicating that eccentricity-oscillations from an undetected giant planet perturber are likely to be suppressed.}
\label{fig:neighbor}
\end{figure*}

Figure \ref{fig:neighbor}a considers a canonical WJ ($m_1 = \mjup$, $a = 0.3$ AU) and fixed giant planet perturber ($m_2 = \mjup$, $\aouteff = 3, 6$ AU) and shows $m_{\rm crit}'$ versus $a'/a$.  A super-earth neighbor ($m' \sim 10 M_{\oplus}$) is extremely effective in suppressing eccentricity oscillations in the WJ, and an Earth-mass neighbor may also prohibit eccentricity oscillations for close separations.

Figure \ref{fig:neighbor}b depicts the sample of WJs with close neighbors from \cite{huang2016} (with the exception of KOI-191.01, since this WJ may actually be solitary [\citealt{law2014}]).  For the neighboring planets in each system, we calculate the value of $\epsilon$, assuming a hypothetical giant perturber $m_2 = 1 \mjup$ and $\aouteff = 5 a$, with $a$ the observed WJ semi-major axis.  Since this sample consists entirely of Kepler objects, many planets lack mass constraints.  WJs without mass estimates have been assigned $m_1 = \mjup$, and the close neighbors have been assigned $m'/M_{\oplus} = 2.69 (R'/R_{\oplus})^{0.93}$ \citep{weiss2014}.  Given these assumptions on planetary masses, the results in Fig.~\ref{fig:neighbor}b should be interpreted with large uncertainties.  Nonetheless, we see that nearly all systems have at least one neighboring planet satisfying $\epsilon \gtrsim 1$, so that eccentricity growth due to the exernal giant perturber is most likely prohibited, or at the very least, reduced.  Given the strong giant planet perturber considered, the values of $\epsilon$ in \ref{fig:neighbor}b represent an optimistic scenario: weaker perturbers will lead to even larger values of $\epsilon$.

We conclude that close (within $\sim [0.1 - 10] a$), low mass ($1-10 M_{\oplus}$) neighbors to WJs are frequently capable of suppressing eccentricity oscillations in WJs.  If eccentric WJs arise primarily due to secular perturbations from distant giant planet perturbers, they should generally lack nearby companions.

\section{Summary \& Discussion}
Many warm Jupiters (WJs) are observed to have exterior giant planet companions.  This paper considers the scenario where WJs form initially with low eccentricities, having reached their observed orbits either through in-situ formation, or disk migration.   In order to produce the modest eccentricities observed in many WJs, we invoke the presence of an exterior giant planet companion that raises the eccentricity of the WJ through secular perturbations.  The eccentricity of the WJ thus oscillates between $e \simeq 0$ and a maximum value $e = \emax$.  In order for the companion to generate the observed WJ eccentricity $\eobs$ through eccentricity oscillations, we require $\emax \geq \eobs$.  Furthermore, the fraction of time spent at eccentricities equal to or greater than the observed value [denoted as $f(e \geq \eobs)$] should be relatively high.  For a WJ with specified properties, these requirements place constraints on the properties of an external companion in terms of its mass, semi-major axis, eccentricity, and inclination. 

In Section 2, we examine the different mechanisms/regimes of eccentricity excitation of a ``canonical'' WJ (with $m_1 = 1 \mjup$, $a = 0.3$ AU) by an outer planetary companion of various masses and orbital properties.  Coplanar and inclined systems are discussed separately, because coplanar systems allow an octupole-level analytic treatment, whereas octupole-level inclined systems require full numerical integrations.  In additional to the secular interactions between the two planets, we also consider apsidal precession of the inner planet due to general relativity and tidal distortion.  For coplanar and moderately inclined systems ($I_0 \lesssim 30^\circ$), the apsidal precession resonance, which occurs when the net precession rates of the two planets (driven by mutual 
interaction and the GR effect) become equal (see Eqs.[\ref{eq:rescondition}]-[\ref{eq:res_low_e}]).  This leads to efficient eccentricity excitation (see Figs.~\ref{fig:emax_coplanar} and \ref{fig:emax_inclined}).  We also show that the extreme eccentricity 
excitation and orbital flip discussed in previous work \citep{li2014} are unlikely to operate for realistic systems (Section 2.3).  For higher mutual inclinations, the Lidov-Kozai eccentricity effect leads to eccentricity excitation.

The main results of Section 2 are summarized in Figs.~\ref{fig:emax_LK} and \ref{fig:frac_LK}. Figure \ref{fig:emax_LK} reveals that coplanar and low-inclination ($I_0 \lesssim 30^\circ$) perturbers may lead to substantial eccentricity increases, with $\emax \simeq 0.2 - 0.6$, where the range in $\emax$ depends on the perturber mass, separation, and eccentricity.  Massive perturbers with high eccentricities are especially effective in producing large $\emax$ over a wide range of separations.  However, despite these large values of $\emax$, the fraction of time the WJ spends in such eccentric states is often small (see Fig.~\ref{fig:frac_LK}).  We conclude that a coplanar or low inclination companion may easily lead to a mildly eccentric WJ (with $e \simeq 0.2$), provided that the perturber is massive and highly eccentric (with $m_2 \gtrsim \mjup$ and $\eout \simeq 0.75$).  On the other hand, such a companion is unlikely to produce a moderately eccentric WJ (with $e \simeq 0.5$), because the fraction of time the WJ spends at or above $e = 0.5$ is very low.

Higher mutual inclinations are generally much more effective in producing eccentric WJs, due to Lidov-Kozai cycles.  Inspecting the high-inclination results (with $I_0 \gtrsim 40^\circ$) in Figs.~\ref{fig:emax_LK} and \ref{fig:frac_LK}, we find that such inclinations may easily produce a mildly eccentric WJ (with $e \simeq 0.2$), since $f(e \geq 0.2) \gtrsim 0.5$ in most cases.  Producing a moderately eccentric WJ (with $e \simeq 0.5$) is also possible, with $f(e \geq 0.5) \simeq 0.3$ for some inclinations.

In Section 3 we apply our method and analysis to observed WJs with exterior giant planet companions.  These systems have measured minimum masses, semi-major axes, and eccentricities for both the WJ and outer planet, but lack information on the mutual orbital inclination \citep[see Table 1 in][for measured system parameters]{antonini2016}.  For each system we have identified the necessary mutual inclinations to produce the observed WJ eccentricity (see Figs.~\ref{fig:wjdata_fiducial}, \ref{fig:wjdata_fiducial_allphase}, and \ref{fig:wjdata_all_experiments}), for several different assumptions of the initial eccentricities and planetary masses of both planets.  The majority of systems require mutual inclinations of at least $40^\circ - 50^\circ$, in agreement with the results of Section 2.  Exceptions are HD159243, HD207832, and depending on the particular assumptions (see Section \ref{sec:additional}), possibly HD169830 and Kepler-432.  The eccentricities of these four WJs may result from coplanar or low inclination configurations under some circumstances (but note the caveat concerning Kepler-432; see the discussion at the end of Section \ref{sec:obsfiducial}).  

Explaining the three most eccentric WJs in the sample (HD74156, HD37605, and HD163607, with $\eobs \gtrsim 0.6$) is more difficult, because we find the fraction of time spent above the observed value is usually less than $20 \%$.  If the eccentricities of these planets are the result of secular eccentricity oscillations from the observed companion, then we are observing them at rather special moments during their oscillation cycles.  On the other hand, such high eccentricities in WJs are also less common, which may help alleviate this issue. 

Since $\sim 50 \%$ of WJs are estimated to have close rocky ``neighbors'' \citep{huang2016},  we have also briefly explored the effects of a third, low-mass planet orbiting close to the WJ (see Section 4).  The precession induced on the orbit of a WJ by such a neighbor may often overcome the precession induced by a more distant giant planet companion, thereby suppresssing eccentricity oscillations.  By comparing the precession rates induced by a low mass neighbor and a distant giant planet perturber, we find that $\sim (1 - 10) M_{\oplus}$ neighbors may frequently suppress eccentricity oscillations in a canonical WJ (see Fig.~\ref{fig:neighbor}a).  We also consider the observed close neighbors to WJs from \cite{huang2016},  and calculate the precession induced in the WJ by the neighbor(s), compared to that due to an undetected giant planet perturber.  We show that even for a strong giant planet perturber, most systems contain at least one neighboring planet likely to suppress eccentricity oscillations (see Fig.~\ref{fig:neighbor}b).   In the \cite{huang2016} sample, four WJs with close neighbors currently have constraints on the WJ eccentricity\footnote{Eccentricities were obtained from exoplanets.org and exoplanet.eu, accessed on July 17, 2017.}, three of which (Kepler-46, Kepler-117, and Kepler-289) have low eccentricities, in the range $\eobs \simeq 0.003 - 0.03$, while the fourth (Kepler-418) has $\eobs \simeq 0.2$.  The fact that WJs with close neighbors tend to have low or modest eccentricities is consistent with our finding that such neighbors probably do not allow the eccentricity of the WJ to grow from secular perturbations from a more distant giant planet.  On the other hand, such lack of eccentric WJs with close neighbors may also simply result from dynamical stability requirements.   

We conclude that the explanation for eccentric WJs proposed in this paper requires that eccentric WJs should generally lack close neighbors of masses $\sim 10 M_{\oplus}$.  The consequence of a lower mass ($\sim 1 M_{\oplus}$) neighbor is less certain, and should be explored in future work via N-body integrations of three planet systems. 

Our results suggest that many observed WJs could have highly inclined ($\gtrsim 40^\circ$) external giant planet companions. This is intriguing, because it requires an initial scattering event to generate the mutual inclination, and therefore the existence of at least three giant planets.  A recent measurement of a high mutual inclination for a WJ with an external companion, using transit-timing and transit duration variations, implies that high inclinations may be relatively common \citep{masuda2017}.  As observations continue to probe mutual inclinations in multiple planet systems (see \citealt{mcarthur2010} and \cite{mills2017} for two examples of mutually inclined systems), a clearer picture of the role of external companions on the eccentricities of inner planets will emerge.  

\section*{Acknowledgments}
We thank the anonymous referee for useful comments.  This work has been supported in part by NASA grants NNX14AG94G and NNX14AP31G, and a Simons Fellowship from the Simons Foundation.  K.R.A. is
supported by the NSF Graduate Research Fellowship Program under Grant
No. DGE-1144153.

{}

\end{document}